**Chapter 12**

# Athermal photoelectronic effects in non-crystalline chalcogenides: Current status and beyond[*]

*Spyros N. Yannopoulos*

*Foundation for Research and Technology Hellas – Institute of Chemical Engineering Sciences (FORTH/ICE-HT), P.O. Box 1414, GR-26504, Rio-Patras, Greece*

## 12.1 Premise and scope of the review

The current critical review aims to be more than a simple summary and reproduction of previously published work. Many comprehensive reviews and collections can be found in the literature [1–4]. The main intention is to provide an account of the progress made in selected aspects of photoinduced phenomena in non-crystalline chalcogenides, presenting the current understanding of the mechanisms underlying such effects. An essential motive for the present review article has been to assess critically published experimental work in the field. There are examples where terminology has been misleading; this has led to classify underrepresented phenomena as new effects.

To make this review self-contained, basic concepts are briefly introduced to render the topic more accessible for non-specialists to assess the current level of understanding of photoinduced phenomena. Then, an epigrammatic overview of the current state of photoinduced phenomena is presented, followed by identification and discussion of the key light-driven athermal phase changes placing the emphasis on "solid"-to-"fluid" (photoplastic) transitions, amorphous-to-amorphous changes, and reversible amorphous-to-crystalline transitions. The last two kinds of transitions are much less explored in comparison to the photoplastic effects. However, the underlying mechanism for all types of transitions mentioned above, which is key to applications, is yet far from being understood either due to misconceptions or because the relevant phenomenology is still not rich enough. No matter how real or exalted the applications of chalcogenides are, the interest in deciphering unique photoinduced phenomena will continue to fascinate scientists.

---

[*] Dedicated to Prof. George N. Papatheodorou on the occasion of his 80th birthday.

As a mature now, and possibly declining field of research in the near future, photoinduced phenomena observed in non-crystalline chalcogenides have on the one hand offered a number of fascinating concepts and ideas but has on the other side being pervaded with weaknesses and frailties. It is acceptable that some degree of speculation would be unavoidable in devising structural models of involved atomic reconfigurations. However, in several cases the proposed structural changes are system-specific, and hence their generalization to any material goes far beyond realism. In addition, the accidental or occasional "deliberate" ignorance of previously published work has led to "*rediscovery*" of known effects, added in the quiver of the photoinduced effects with different terminology. In some cases, this has triggered a blurred scenery, which can rather easily again lead to misconceptions and fallacies. Finally, despite some other categories of materials (liquid crystalline polymers etc.) exhibit a dazzling similarity in photoinduced effects, the field of amorphous chalcogenides has demonstrated remarkable introversion with negligible cross-talk between the two disciplines. Only very recently some parallelisms have been envisaged.

**12.2   Introductory remarks**

The term *chalcogen* was historically coined unofficially by W. Fischer, back in 1932, to describe the atoms of group 16 in the periodic table, i.e. O, S, Se, Te, and Po [5]. This composite term derives from the Greek words chalkos ($\chi\alpha\lambda\kappa\acute{o}\varsigma$) and genos ($\gamma\acute{\varepsilon}\nu o\varsigma$) meaning copper and gender, respectively, which at a first glance seems irrelevant for the above purpose. It was only after 1941, when IUPAC officially adopted this term in the nomenclature of Inorganic Chemistry, when the term chalcogen became widely popular as it seemed analogous to the term halogen. Despite that oxygen compounds are traditionally called oxides, literally, all compounds containing at least one of the chalcogen atoms are termed as chalcogenides. Alloying chalcogens with elements such as pnictogens (group 15; P, As, Sb), tetrels or tathogens (group 14; Si, Ge) and triels (group 13; Ga, In, Tl) non-crystalline chalcogenides - either as bulk glasses or in the form of amorphous thin films - can be prepared in a number of preparation techniques [6]. Strong interest for another family of chalcogen-based layered compounds, the so-called transition metal dichalcogenides of the form $MX_2$ (M: Mo, W, Ta, etc.; X: S, Se,

Te) has recently resurfaced to the frontline of materials science owing to new methods for their growth in mono- and few-layer 2D crystals exhibiting novel physics compared to their bulk counterparts [7].

Non-crystalline chalcogenides continue to attract the attention of research community, throughout more than sixty years after their discovery [8], for both scientific and technological reasons. One of the first striking findings was that chalcogenide glasses exhibited electric conductivity as high as $10^{-3}$ $\Omega^{-1}$ cm$^{-1}$, i.e. almost five orders of magnitude higher than that of oxide glasses, which is of a purely electronic origin. Their semiconducting nature was initially challenged by theoretical solid state physicists [9] who supported that without the lack of long range order (crystal lattice) a three dimensional system could not preserve a bandgap. A rigorous theoretical proof for the existence of a bandgap in amorphous semiconductors was undertaken by Weaire and Thorpe [10]. Exploiting their semiconducting properties and the moderate bandgap exhibited by these materials, illumination by visible light with energy comparable to the bandgap energy can create electron-hole pairs. The fate of the photoexcited carriers determines the response of the material to light: (a) electrical response (separation), (b) photoluminescence (radiative recombination), and (c) atomic arrangement change (non-radiative recombination). The last case, being the origin of photoinduced structural (PiS) changes, is perhaps the most intriguing one. Impurity content, temperature, and the concentration and charge state of native defects are factors that control the recombination of the photoexcited carriers.

PiS changes are the hallmark of non-crystalline chalcogenides underlying a vast number of changes in physical/chemical properties, known as photoinduced phenomena. The term photoinduced throughout this chapter adopts the purely *athermal nature* of the effects where no energy is transferred as heat from the irradiation source to the material. Although the studies of photoinduced effects abound in the current archival literature, a significant fraction of *putative* photostructural changes is essentially – either partly or solely – thermally induced.

The PiS changes give rise to hierarchical changes in the short and intermediate range structural order. Proliferating from the local "molecular" level to macroscopic scales; hence affecting macroscopically observed physical/chemical properties such as density, mechanical properties (hardness, elastic constants), rheological properties (viscosity, the glass transition temperature), optical and electronic transport properties as well as the capacity of chemical solubility. Applications of amorphous chalcogenides exploit either their transparency to infrared light (passive) or their responsiveness to near-bandgap illumination (active). The latter, i.e. the external-stimuli control of the structure and properties of chalcogenides is perhaps scientifically the most interesting case in view of the challenges it poses for a microscopic understanding of photoinduced phenomena.

## 12.3  Basic concepts and definitions

Lacking all the beneficial consequences derived from the concept of lattice in crystals, such as Brillouin zones, Bloch states and so on, amorphous materials are devoid of a corresponding rigorous mathematical description. Although the lack of these *mathematical* tools debilitated our efforts and stymied for long our understanding of relevant phenomena, other properties accompany non-crystalline media endowing them unique properties. These new features, arising from the loss of long range periodicity, promote structural changes induced by light, rendering amorphous solids, arguably, more interesting than their crystalline counterparts. Band tailing and localization of photoexcited carriers are two such attributes arising from disorder, affecting the fate of the carriers. This will further affect the probability for changes in atom arrangements and the valency of the atoms.

The electronic structure of the outer shell of chalcogens contains six electrons; two of them are paired in an *s* state and four electrons are in the *p* state. Two out of the *p* electrons are paired. The electron configuration reads as $ns^2 p_x^1 p_y^1 p_z^2$. Therefore, chalcogens form two bonds via the unpaired electrons of the $p_x$ and $p_y$ orbitals, while the lone-pair (LP) of electrons in the $p_z$ orbital plays a vital role in PiS changes. The electronic structure of chalcogen atoms and chalcogen solids is illustrated in Fig. 1. The *s*, σ and LP states form the valence band (VB), with the LP being at the top of the VB, while the *p* orbitals form the conduction band (CB). The theory of electronic structure in non-crystalline chalcogenides was primarily formulated by Mott [11] and Cohen *et al*. [12]. Mott introduced the concept of *mobility edge* which defines a critical density of states in the amorphous solid below which all states are localized, but above which free carriers exhibit a finite mobility for all states. In contrast to Bloch states in crystalline solids, which describe the extended nature of the electron wavefunction, the electronic states in amorphous solids may be characterized by *localized* states. Quantum mechanical tunneling between localized states could possibly lead to apparently extended states, as far as transport is concerned. The energy separation between the mobility edges of the valence and conduction bands defines the *mobility gap*. By conception, the model on electronic structure developed by Mott [11] and Cohen *et al*. [12] tacitly assumed that all atoms are normally coordinated obeying the 8-N rule (N≥4), where N is the number of valence electrons. Hence, localized states in the gap are caused by disorder whose origin arises predominantly from bond length and angle deviations from the norm and possibly also comes from wrong bonds between atoms. However, structural studies have shown that atoms with coordination deviating from the 8-N rule constitute an appreciable fraction in amorphous solids. These atoms and their immediate environment are called *defects*. It should be noted that defects in crystals are primarily imperfections of the lattice, appearing as dislocations, vacancies, interstitials, and so on, which obviously have no meaning in amorphous solids. The coordination number, the charge of each atom and the type of neighbors are parameters used to account for defects in disordered solids. A schematic representation of the above description is shown in Fig. 2.

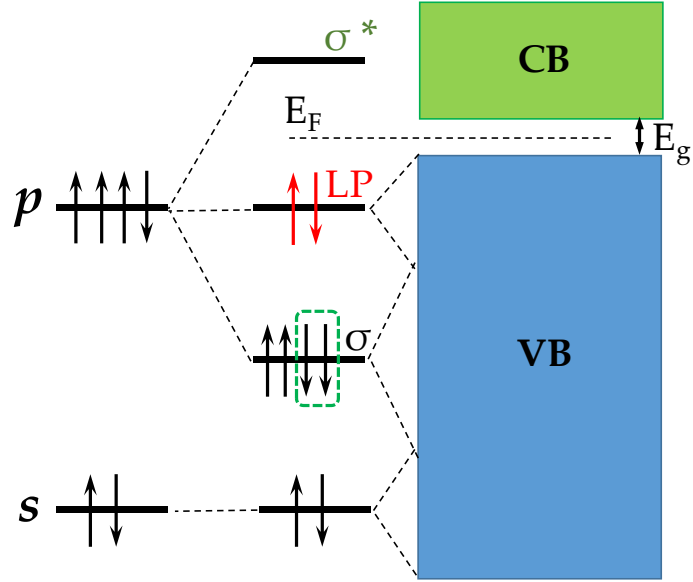

**Fig. 1:** Electronic structure of an isolated chalcogen atom (left) and its evolution when two such atoms bond together. A new configuration appears (middle) where two extra electrons come from the neighboring atom. The σ, σ* and the LP energy states split as designated. These states transform to bands in the solid (right).

Cohen *et al*. [12] proposed that for solids with high degree of disorder, the valence and conduction band tails, which describe the localized states, could be extended to such range that overlap in the center of the gap, providing a finite density of states at the Fermi energy ($E_F$) and hence causing pinning of the $E_F$. In this context, the material could be considered as a semiconductor if the density of states at $E_F$ is less than the critical value which defines the mobility edge. Otherwise, the solid would be an amorphous metal. To account for the apparent contradiction between the pinning of the $E_F$ and the lack of unpaired electrons in amorphous chalcogenides, as no electron spin resonance signal is observed, Anderson [13] suggested a mechanism through which almost all spins can become paired and hence negatively charged (defects) upon the adoption of an extra electron. This process increases the energy due to Coulombic interactions among electrons by an amount known as the Hubbard energy U. The low coordination number of chalcogen atoms brings in structural flexibility that enhances the possibility of a strong electron–lattice interaction. The energy gain resulting from this interaction, as a consequence of disorder, (i.e. the distortion of the chemical bond) overcompensates the repulsion developed within each electron pair. Thus, the effective Hubbard energy becomes negative, at sufficiently strong electron-lattice interaction, and the charged defect center atomic configuration is more stable than the neutral one.

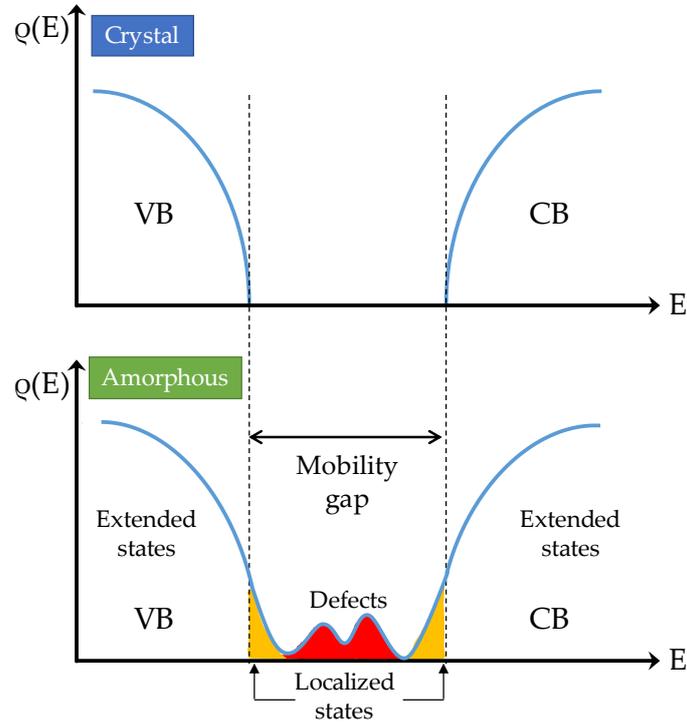

**Fig. 2:**   Schematic diagram of the density of states in crystalline (up) and amorphous solids (down).

The following notation will be used $X_n^q$, where $X$ denotes the type of atom, $n$ is the coordination number and $q$ symbolizes the charge of the atom under consideration. Representing C, P and T atoms from groups 16, 15 and 14, respectively, then the normal bonding state is provided by the configurations $C_2^0$, $P_3^0$, and $T_4^0$. Deviations from these configurations imply the formation of defects.

Typical defect pairs with low formation energy are $C_3^+ - C_1^-$, $C_3^+ - P_2^-$, $C_3^+ - T_3^-$, etc. A schematic of their structure is shown in Fig. 3. The overcoordinated and undercoordinated atoms are positively and negatively charged, respectively. Their creation is the result of the existence of the electron lone pair of the chalcogen atoms. Removing one electron from a chalcogen atom creates a positively charged ion $C^+$ which can form an additional bond. The electron that has been removed undermines an existing bond to an atom it is attached. The average coordination number is preserved in each such pair. Atoms involved in the above mentioned mechanism are called valence alternation pairs (VAPs). The rather low creation energy of VAPs, which refers to the Coulomb energy (resulting from the presence of the extra electron on the negatively charged center) implies a density of $10^{18}$ – $10^{20}$ VAPs cm$^{-3}$ frozen-in at the glass transition temperature. The VAP defects shown in Fig. 3 have transient character and the dissipation of the energy during the non-radiative recombination takes place via structural deformation mediated by phonon emission. It is highly probable that the original configuration will be restored but other atomic arrangements may arise as well with high likelihood. In this last case, the structure is severely modified as bond interchange involves atomic motions reminiscent to diffusion.

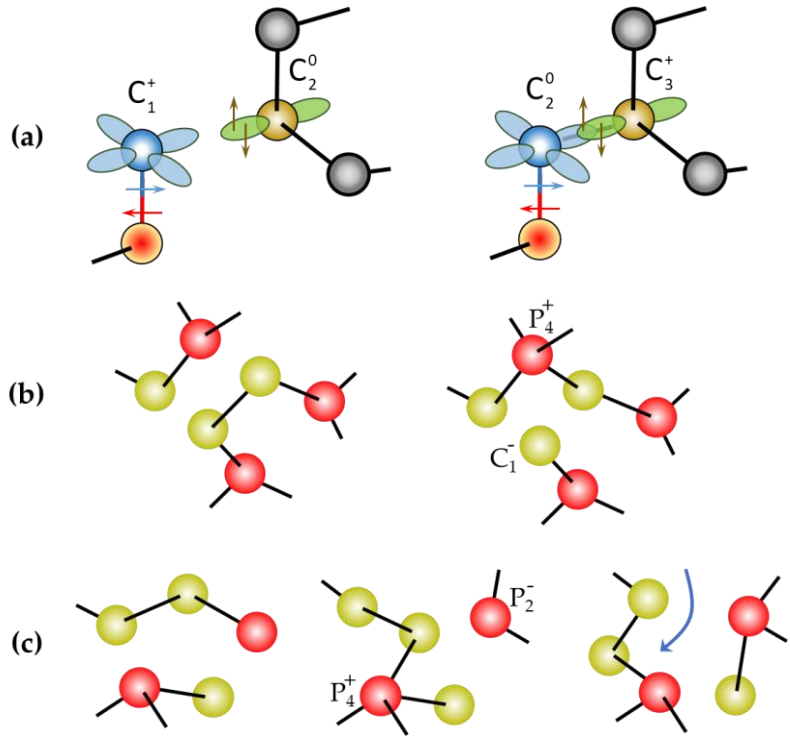

**Fig. 3:** Schematic illustration of various VAPs configurations. **(a)** the formation of a VAP in elemental Se where a positively charged dangling bond $C_1^+$ bonds to a two-fold (normal) coordinated $C_2^0$ atoms to form an overcoordinated $C_3^+$ defect (adapted from Ref. [14]). **(b)** A normal coordination (left) of pnictogen-chalogen system transforms to an intimate VAP (left) as a self-trapped exciton. (c) PiS changes involving bond reorientation mediated via a self-trapped exciton.

The idea of a strong electron-phonon coupling at defects, mentioned above, is central to understanding localized states in non-crystalline chalcogenides [15,16]. The strong phonon coupling is the result of the energy released when an atom (defect) adopts a bonding rearrangement to a local coordination different from the bulk and provides a plausible mechanism to explain the faster rate of non-radiative release of excitation energy in comparison to the radiative process [17]. Based on these arguments, Street envisaged a mechanism for the fast non-radiative recombination, according to which the photoexcited electron-hole pair forms local transient bonding arrangement [18]. The latter consists of a *self-trapped exciton* (Fig. 4) or else a coordination defect pair, i.e. the VAP. In general, the creation of a defect pair reaction can be written as:

$$A_m^0 + B_n^0 \leftrightarrow A_{m+1}^+ + B_{n-1}^- \qquad (12.1)$$

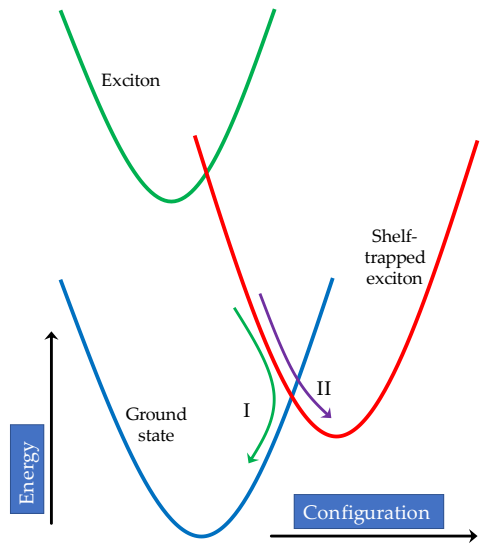

**Fig. 4:** Configuration coordinate diagram showing two possible recombination paths after photoexcitation in amorphous chalcogenides. (I) Direct recombination to the ground state. (II) Creation of a metastable state (VAP) in a potential well that requires energy (heat) to restore the equilibrium state. Based on Ref. 18. Copyright © 1977 Published by Elsevier Ltd.).

Charge transfer from an atom to another lowers the activation energy for the defect creation. When the two VAPs are close enough forming an *intimate* VAP, they could self-annihilate, as is described by the back arrow in Eq. (1). For example, in the case of a pnictogen-chalcogen system, the above concept for the formation and annihilation of the transient defect state can be expressed as $P_3^0 + C_2^0 \rightarrow P_4^+ + C_1^- \rightarrow P_3^0 + C_2^0$. A schematic configuration-coordinate diagram, proposed by Street, showing how a transient bonding arrangement can dissipate the recombination energy is illustrated in Fig. 4. The rich variety of local bonding environments in amorphous solids and the structural flexibility which configurations with low steric hindrance offer, renders the interconversion of such defects possible via the reaction (2),

$$A_{m+1}^+ + B_{n-1}^- \leftrightarrow A_{m-1}^- + B_{n+1}^+ \qquad (12.2)$$

VAP pairs can capture two carriers of the appropriate charge thus reverting its charge. This valence change will alter the coordination number, rendering e.g. the initially positive overcoordinated defect A to a negative undercoordinated state.

## 12.4. A general perspective on PiS changes: unique to non-crystalline chalcogenides or universal property?

Because the very first photoinduced effects explored in non-crystalline chalcogenides such as photodarkening and photodissolution of metals (photodoping) were shown to be extremely suppressed in their crystalline counterparts, the general perception that prevailed was that this is an intrinsic merit of non-crystalline semiconductors. Although not widely known, PiS changes are not unique to non-crystalline chalcogenides. In certain cases, where light illumination causes electronic excitations with sufficiently long lifetimes, observable effects on various types of phase transitions follow. Typical examples of such transitions include structural transformations, nonmetal-to-metal transitions, changes related to supercooling, supersaturation, and so on [19]. A general conception that is frequently adopted in such non-equilibrium process is depicted in Fig. 5. The potential energy surface (PES) representing the Gibbs potential, separates two states or two phases A and B by a high barrier that can be overcome by thermally activated processes, when the system is in the ground state. Electron excitation brings about changes in bonding and the new atomic configuration may be described by a lower barrier height in the non-equilibrium state. This scenario presupposes that the excited electrons have a long enough lifetime allowing the system to respond via structural changes. Typical phase transitions in the above context include athermal melting observed experimentally by ultrafast time-resolved X-ray diffraction techniques, in an amorphous semiconductor InSb [20] and single-crystal Ge films [21], showing that non-thermal or electronic melting of materials might be a universal property. Photoinduced bond weakening is the origin of athermal photomelting, while optically-induced bond strengthening can cause effects such as non-thermal vapor condensation or transitions of liquids into solids [22]. A more elaborate discussion on photoinduced phase transition can be found elsewhere [23].

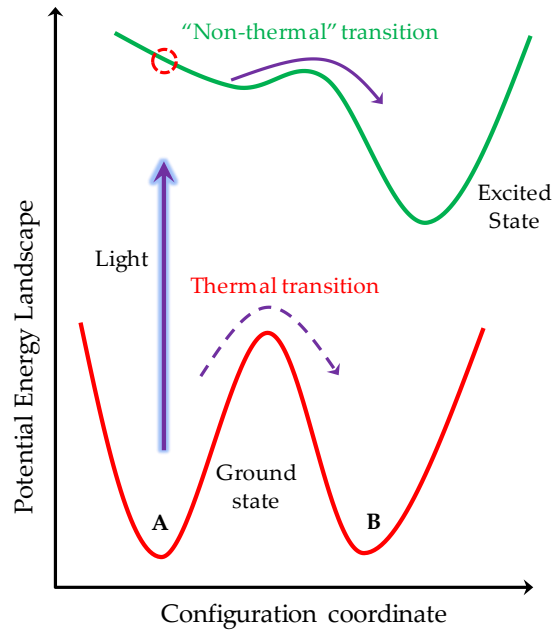

**Fig. 5:**   Configuration coordinate diagram showing a path of photoinduced changes of the potential energy surface from state A to state B. The much smaller energy barrier in the excited nonequilibrium state in comparison to the ground state, implies practically an athermal transition as the ambient temperature could provide the necessary energy for the transition with no additional external temperature rise. Adapted from Ref. 19.

Another well-explored case of materials undergoing interesting PiS changes are polymers containing azo groups (-N=N-). These materials have been intensively studied as they exhibit remarkably similar effects under light illumination with chalcogenides [24]. The *trans-cis* photoisomerization is the photochemical reaction standing as the basis of the effects in azo-polymers. Notably, the same terminology is being used for azo-based polymers, while the two research sectors, i.e. chalcogenides and azo-polymers have progressed in parallel over the years with almost no interaction.

## 12.5 Types of photoinduced phenomena in non-crystalline chalcogenides and their classification

A large number of photoinduced phenomena, straddling a wide gamut of related physicochemical properties, has been reported up until now for non-crystalline chalcogenides. These phenomena involve elemental chalcogens, binary, ternary and more complex systems. However, as will be discussed in detail in the next sections, photoinduced changes appear more prominent at certain compositions, since local structural motifs can facilitate atomic rearrangement. The underlying light-induced changes range from relatively subtle ones concerning minimal atomic displacements to more extensive atomic and molecular rearrangements including the formation of new species (i.e. molecular units formed in network-like structures). The proliferation of these structural reconfigurations causes a variety of physical and chemical changes. We should stress that although in several descriptions, photoinduced structural changes are treated separately from photoinduced physicochemical changes (observed as a change in some observable property), this distinction is pointless as changes in the structure (i.e. atomic configuration) are ultimately the origin of physicochemical modifications. Alternatively, when structure changes, properties also vary.

Photoinduced phenomena can be categorized into various classes based primarily on their magnitude and the reversibility pathways they exhibit following cessation of the irradiation. In this sense, modifications can be:

(i) transient; persisting only as long as illumination lasts ("*light on*" conditions)

(ii) partially reversible; the illuminated volume attains partly its original properties after applying an external stimulus, most frequently heat but also light in rare cases

(iii) fully reversible; the application of the external stimulus reverts the illuminated volume back to its initial state

(iv) irreversible or permanent; the modifications in structure cannot be eliminated by any means of post treatment

(v) a mixture of transient and reversible/irreversible; essentially, it is plausible to consider that all photoinduced effects bear a transient character, which depends upon the type of photostructural changes involved. The extent of the transient change in relation to the residual one after light cessation can be realized only if measurements are conducted *in situ*.

On another front, the changes can be *scalar* or isotropic and *vectoral* or anisotropic denoting the influence or not of the orientation of the light polarization, respectively. Several review articles and chapters have compiled the various photoinduced phenomena proving details about their origin and the existing controversies for the nature of such effects, which will not be discussed here in depth [2, 3, 25–28].

Besides the presence of LP electrons and their role in promoting photoinduced phenomena in non-crystalline chalcogenides, another crucial reason for facilitating PiS changes is the possibility to form homonuclear or so called "wrong" bonds between cation atoms in stoichiometric compounds. Indeed, such bonds can readily be created (by light) and annihilated (by heat) as the bond formation of such atoms (M: metals and metalloids of groups 14 and 15) have formation energies in the range $150 - 220$ kJ mol$^{-1}$, which is comparable to heteronuclear bonds between these and chalcogen atoms (C). In general, the following reaction describes usual PiS changes:

$$2|M-C| \xleftrightarrow{h\nu,\ T} |M-M|+|C-C| \qquad (12.3)$$

Typically, illumination shifts the equilibrium to the right creating chemical disorder, whilst annealing restores chemical order by shifting the equilibrium to the left.

Well-annealed films and bulk glasses prepared by melt-quenching (which are fairly well-annealed due to their slow cooling) exhibit predominantly reversible effects. On the contrary, as-prepared and poorly-annealed films prepared mainly by thermal evaporation and ultrafast quenched glasses exhibit certain irreversible changes which are primarily morphological, structural or compositional. The differential evaporation rate of various elements and/or species is the main cause leading to films with different structure in comparison to the bulk glass. An early Raman study (Fig. 6, *left*) on the irreversible thermo- and photo-structural transformations in As$_2$S$_3$ films has been reported by Solin and Papatheodorou [29], assigning the observed changes to photoinduced polymerization of metastable molecular units formed during the deposition process.

Further Raman studies on the thermo- and photo-structural transformations in As$_2$S$_3$ films were reported by Frumar et al [30]. (Fig. 6, right). The Raman spectra show that partial photo-polymerization takes place in the course of irradiation, as sharp Raman bands indicative of the molecular species are detected. Annealing causes more extensive polymerization, re-attaining the chemical order which is observed in the bulk glass. Sharp bands of As-As bonds at energy lower than 250 cm$^{-1}$ and S-S bonds at ~ 490 cm$^{-1}$ are evident supporting the equilibrium reaction (3), which assumes the form:

$$2|As_2S_3| \leftrightarrow |As_4S_4| + S_2 \qquad (12.4)$$

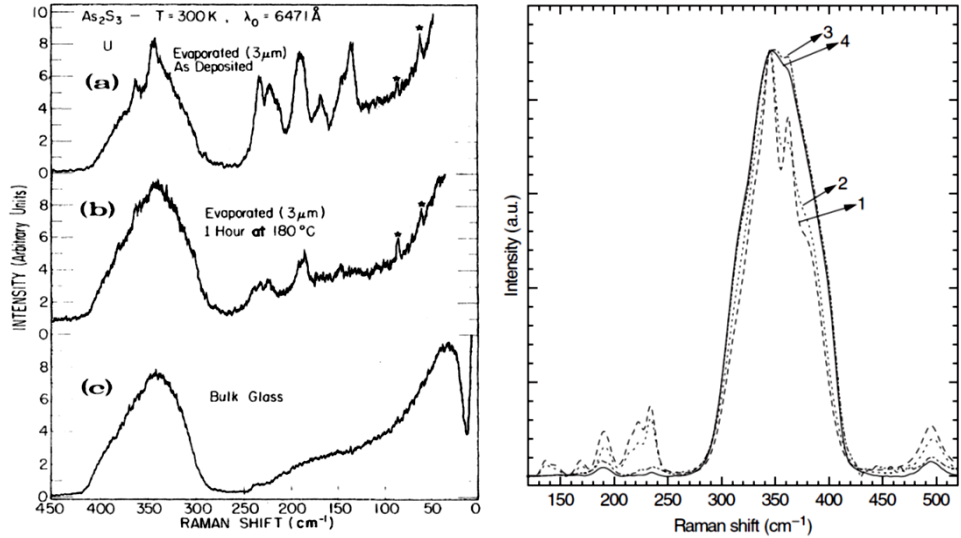

**Fig. 6:**   *Left*: Raman spectra of $As_2S_3$ (a) an evaporated as-deposited film, (b) annealed at 180 °C for 1 h, (c) bulk glass. Reprinted (Figure 1) with permission from Ref. 29, S. A. Solin and G. N. Papatheodorou, Phys. Rev. B 15, 2084 (1977) by the American Physical Society. *Right*: Reduced Raman spectra of $As_2S_3$ thin films and bulk glass. 1: as-evaporated, 2: exposed, 3: annealed, 4: bulk glass. Reprinted with permission from Ref. 30.

A more detailed combined resonance and off-resonance Raman study of the photo- and thermo-structural changes in bulk glassy $As_2S_3$ (studied under vacuum) was reported by Yannopoulos *et al.* [31]. Bandgap illumination was found to induce much stronger structural changes than ever reported for this bulk glass, which was attributed to the absence of oxygen in the course of illumination. The formation of As-As bonds (band at ~234 cm$^{-1}$) is the main light-induced structural change (Fig. 7, *left*). This observation, together with the absence of S-S bond formation, points to the failure of models adopting the formation of realgar-type structures via the frequently adopted reaction (12.4) and calls for new structural defects based on centers with an increased coordination number. The kinetics of photostructural changes exhibits moderate departure from the single exponential behavior and the characteristic time scale is of the order of several minutes. This time scale was found to be comparable to the characteristic scale of the time dependence of photodarkening for the same glass [32].

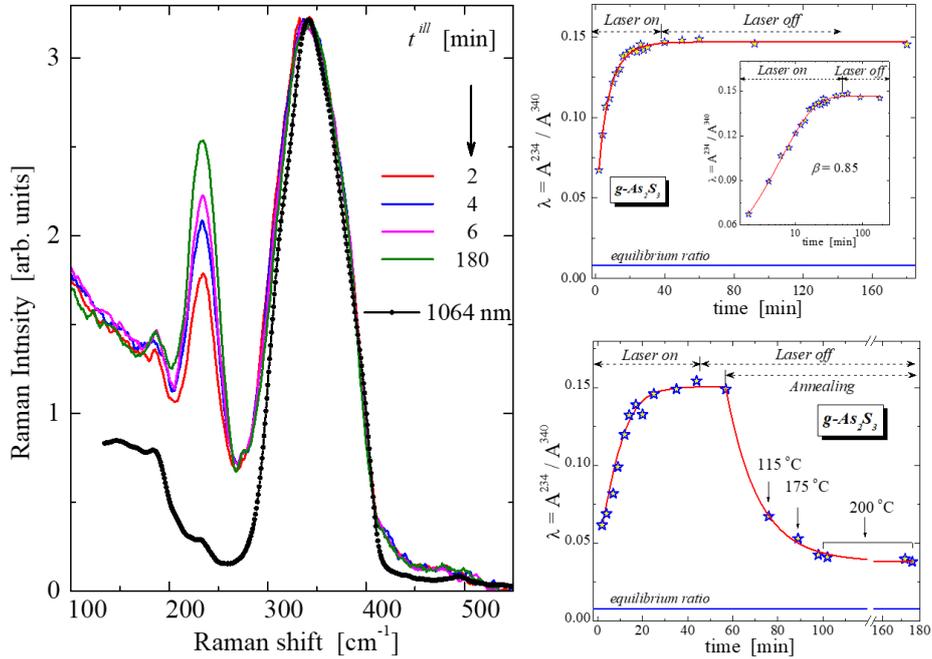

**Fig. 7:** *Left*: Raman spectra of glassy As$_2$S$_3$, excited by bandgap light, at various illumination times as shown in the legend. Solid-dots lines represent the Raman spectrum of the glass at 25 °C recorded at conditions very far from resonance (1064 nm, 1.17 eV). The spectra have been normalized at the intensity maximum near 340 cm$^{-1}$ *Right*: Time dependence of the peak area ratio $\lambda(t) = A^{234}/A^{340}$ of glassy As$_2$S$_3$ under bandgap irradiation. The peak at 234 cm$^{-1}$ arises from photo-induced As-As wrong bond, while the peak at 340 cm$^{-1}$ denotes the As-S vibrational mode of the AsS$_3$ pyramidal unit. The horizontal solid line at $\lambda = 0.008$ corresponds to the value obtained under non-resonant Raman scattering (1064 nm). Reprinted from Ref. 31. Copyright © 2012, John Wiley and Sons.

Summarizing the various states that non-crystalline chalcogenides can adopt under the action of the two main external stimuli, i.e. heat and light, Fig. 8 illustrates the interrelationships among them. The structure of as-deposited films and ultrafast quenched glasses contains an appreciable fraction of structural defects, trapped-in during the preparation procedure. These defects tend to increase the degree of disorder. Reversible transformations can take place between the annealed and irradiated states. The right panels in Fig. 8 show modifications in the coordination number and the mean square relative displacement when amorphous Se is subjected to various treatments [33]. Under irradiation, the mean square relative displacement is higher than the post-irradiated state, denoting the partially transient character of the effect. Appreciable transient photostructural changes measured upon illumination take place for $As_{50}Se_{50}$ films prepared by pulsed laser deposition; this observation will be discussed in detail in Section. 12.8.1.

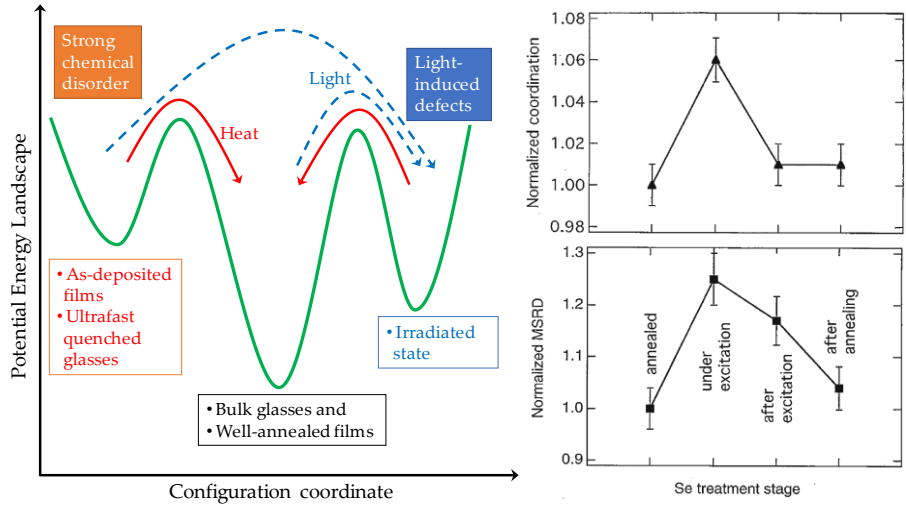

**Fig. 8:** <u>Left</u>: Schematic representation of the free energy vs. configuration; the depth of each minimum denotes the stability of the state. Single and double arrows symbolize irreversible and reversible processes, respectively. Solid- and dashed-line arrows denote light and heat processes, respectively. <u>Right</u>: Changes in the coordination number (*up*) and the means square relative displacement (*down*) of amorphous Se at various treatment steps. Reprinted (figure 9) with permission from Ref. 33, [A.V. Kolobov, H. Oyanagi, Ka. Tanaka, K. Tanaka, Phys. Rev. B 55, 726 (1997)] by the American Physical Society.

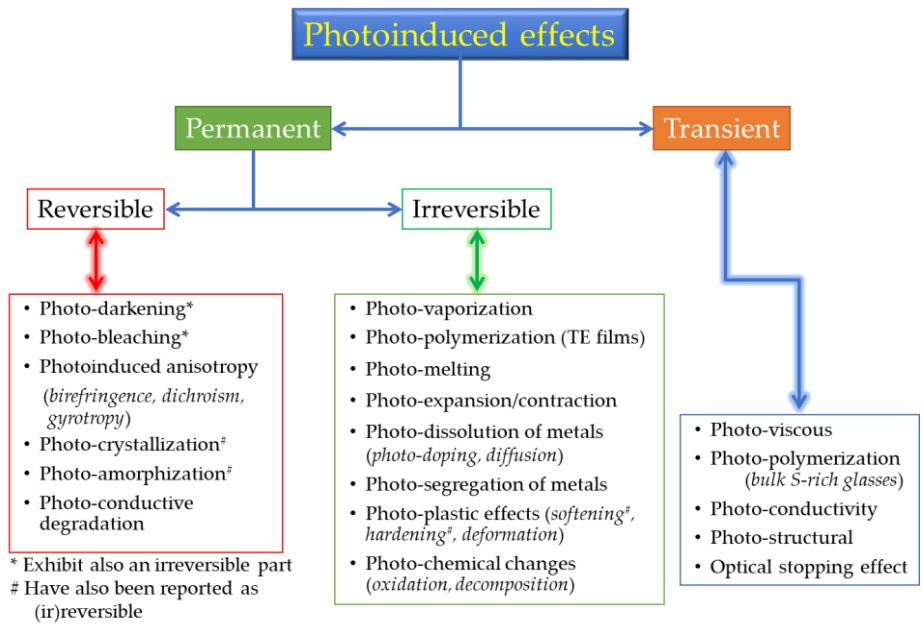

**Fig. 9:** Synopsis of the classification of photoinduced effects.

A general classification of the various photoinduced effects which have been explored and reported in the literature is compiled in the diagram shown in Fig. 9. In certain cases, the effects have been exhaustively studied for decades, while in other instances there are only a few reports. However, in both cases, the structural models which have been developed to account for the origin of these effects appear in certain cases to be inconclusive or self-contradictory, as they fail to clarify all aspects of the phenomenon under question. Despite that certain photoinduced effects may have common origin, different terminology has been coined to express the same effect, adding confusion. This is common for the photoplastic effects which are described in detailed in Section 12.6. Finally, it should be stressed that most of these effects have some transient character, while some other may exhibit a reversible and an irreversible part as well.

## 12.6 Unified perspective on several photoinduced phenomena underpinned by the photoviscous effect

Several of the photoinduced phenomena described in the previous section, especially those related to optical properties, have been thoroughly studied and are now considered as sufficiently understood. In contrast, the situation concerning photoinduced changes in mechanical and rheological properties is not yet clear. The main reason is that no systematic attempt has been made to unify the effects which appear as ramifications of the underlying cause of such effects, i.e. the *athermal* photoinduced lowering of viscosity under illumination, the so-called *photoviscous effect*. This effect pertains to the photoinduced transformation of the illuminated volume from a solid glassy state to a highly viscous "fluid" or alternatively to a transition from a *brittle* state to a *plastic* one.

A detailed account of the confluence of such effects has been provided by Yannopoulos and Trunov [34], who suggested that several phenomena following the photo-viscous effect may bear a common origin. In particular, there has been confusion on the terminology and misconception in the nature of various photoinduced effects in the literature concerning mechanical and rheological properties, such as elastic constants, microhardness, stress relaxation, viscosity, glass transition temperature ($T_g$), mass transport, surface morphology and so on. Therefore, miscellaneous changes observed under illumination of the above-mentioned properties have been called in a number of ways including terms such as photodeformation, photohardening, photosoftening, photofluidity, photoinduced structure (stress) relaxation, photoinduced mass transport, photoinduced relief grating, photomelting and optomechanical effect. It should be noted that slight variations of the experimental parameters, e.g. size of the sample, adherence to the substrate (i.e. free-standing or mounted), duration of illumination, etc. can result into qualitatively different observations, albeit the underlying phenomenon might the same. This has led to exaggeration in claiming the invention of new *putative* photoinduced effects.

In their pioneering works, Nemilov and Tangatsev [35–38] coined the term photoviscous effect and explored in detail the athermal decrease of viscosity upon irradiation. It is essential to stress that increasing the material's fluidity under irradiation, as a result of bond rearrangement, renders the material amenable to mechanical modifications and morphological transformations. These changes can be observed either as shape changes under the application external forces or as relief gratings formed as the result of mass transport. For clarification, if only illumination by light is involved the term "photoviscous" seems more appropriate, whilst in the combined action of light and an external mechanical stimulus the term "photoplastic" appears more suitable to account for the emerging morphic effects. Notably, even the above distinction might not be entirely clear since morphic effects, such as surface relief deformations, can appear in the absence of an external mechanical stimulus.

### *12.6.1  A brief timeline of photoviscous and photoplastic effects*

#### *12.6.1.1  Early studies: photoplastic effects in crystalline semiconductors*

Perhaps, the first report on photoplastic effects appeared almost a century ago by Vonwiller on the photoinduced changes of the elastic properties of glassy Se [39]. Upon illumination, the material exhibited enhanced yielding compared to the "dark" behavior under the application of distorting forces. Similar effect was observed for crystalline Se albeit of much lower magnitude in relation to the glass.

A question that resurfaces over the years is whether photoinduced effects are unique to non-crystalline or aperiodic solids or if they are also observed for crystals. The general perception is that crystals exhibit much less ability to be *molded* by light owing to the restrictions imposed by lattice periodicity, as the atoms must lie on certain positions. This constraint limits the various scenarios offered for the recombination details of electron-hole pairs. The effect of disorder-induced localization of electron and hole states at band edges avails amorphous materials offering more possibilities for the photogenerated electron-hole pairs. However, although structural inflexibility renders crystalline solids less amenable to photoinduced effects in comparison to amorphous solids, photoplastic effects in crystals were reported several decades ago. These effects concern the athermal light-induced plasticity in crystalline semiconductors and have been explored since 1957 for crystalline Ge, Si and InSb [40]. Even at such early times, the term *photomechanical effect* was coined for the observed effect, since it refereed to the considerable hardness (flow stress) increase of the illuminated crystal surface, triggered by infrared and ultraviolet light. This is usually mentioned as the positive photoplastic effect (photohardening), while for some crystals the opposite behavior was observed (negative photoplastic effect, photosoftening) [41].

Illumination power, temperature, and light wavelength are parameters that affect the photomechanical effect. In particular, the effect saturates at high photon fluxes, while it exhibits unexpected (anomalous) behavior, i.e. photoinduced hardening decreases at elevated temperature. Finally, the effect demonstrates strong spectral dependence, being maximized for light energy comparable to the bandgap of the crystal. The explanation of the photomechanical effect was based on the idea that the electron distribution within a dislocation affects strongly the energy of dislocation. Thus, the redistribution of electrons upon illuminating a semiconductor exerts a significant effect on the mechanical properties, through the increased mobility of dislocations due to photoinduced increase of carrier concentration. First attempts to account for the photohardening effect were based on the modification of the motion of dislocations during illumination due to the change of free electrons interacting with the moving dislocations [42–44]. It was also postulated that the photoexcited holes are trapped, which can generate doubly charged ions. These ions can interact strongly with dislocations, thus enhancing flow stress by dislocation locking [45,46]. Such effects have been documented for several single crystals [47], and were assigned to the decrease of microhardness due to chemical bond weakening and isotropization of the crystal caused by the "photogenerated antibonding quasiparticles".

### *12.6.1.2 Photoinduced deformation in amorphous chalcogenides: Conversion of light energy to mechanical energy*

Matsuda *et al*. reported perhaps the first reversible deformations observed in amorphous chalcogenide films, i.e. $As_{20}S_{80}$ and $Sb_2S_3$ deposited on mica substrates under bandgap light illumination, which were attributed to photoinduced structure relaxation [48,49]. Thermal expansion processes, due to absorption of the incident light, were suggested to play a dominant role in this effect, as the temperature rise due to illumination affects structural relaxation. In parallel, a larger number of chalcogenides, members of the As-Ge family were explored by Igo *et al* [50], while details on photoinduced stress relaxation of a-Se/mica film bilayer was reported by Koseki *et al*. [51] The configurational coordinate diagram was invoked to account for the photoinduced stress relaxation which was considered as the result of non-radiative recombination processes. Rykov and coworkers [52,53] devised a more elaborate method to study in detail the photoinduced deformation in amorphous films (Se) and bulk glasses ($As_2S_3$). Polarization dependent reversible contractions and dilations of a-$As_{50}Se_{50}$ deposited on a cantilever's surface have also been reported [54]. Some other aspects of scalar and vector photodeformations in amorphous chalcogenides are described elsewhere [55].

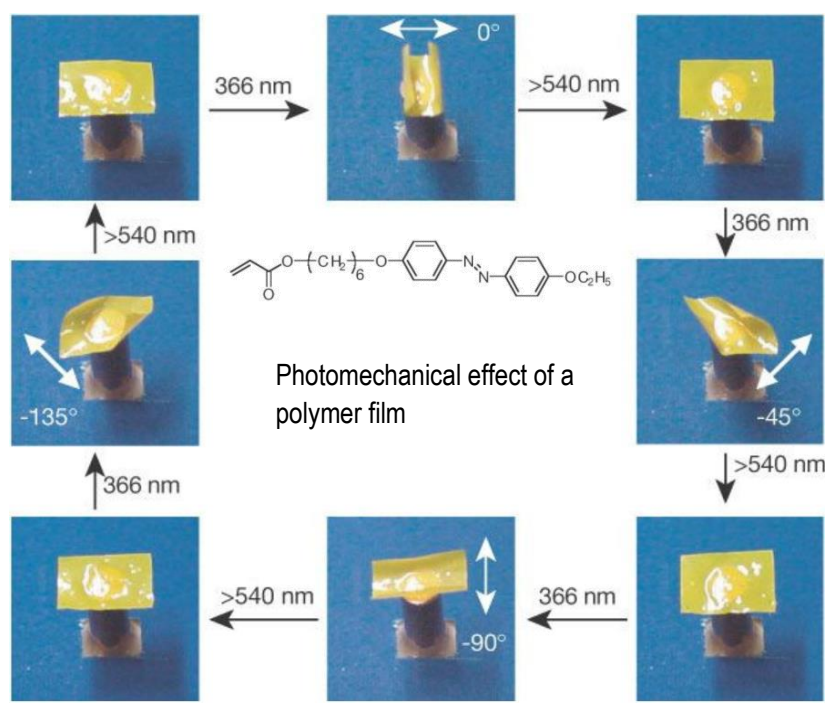

**Fig. 10.** Controlling the bending direction of a film by linearly polarized light. UV light (white arrows) at 366 nm induce bending, while visible light (>540 nm) flattens the film. Film dimensions: 4.5×3 $mm^2$, thickness 7 μm. The deformation and flattening time scales are both of ca. 10 s. Reprinted from Ref. 57. Copyright © 2003, Springer Nature.

It would be instructive to mention that photoinduced deformations are very common in azobenzene containing materials. Light enables the interconversion between the two geometrical isomers and the cumulative effect becomes evident as macroscopic mass transport. Typically, illumination triggers *trans→cis* isomerization whilst the opposite transformation takes place either by light of heat [56, 57]. Impressive reversible light-induced deformations observed on mm-scale films of a liquid-crystal is shown in Fig. 10. Analogous studies have failed to demonstrate so nicely and in a controlled way photoinduced deformation (light-induced actuation) in in amorphous chalcogenides [55].

### *12.6.2   The photoviscous effect in bulk chalcogenide glasses*

Nemilov and Tagantsev undertook the most systematic exploration to account for the origin of light-induced morphic effects [35-38], which will be surveyed in the current Section. They followed an approach enabling them to conduct *in situ* measurements of viscosity changes under illumination for a number of bulk chalcogenide glasses. The light-induced, athermal decrease of the material's viscosity under equilibrium conditions was described as the photoviscous effect. In their series of experiments Nemilov and Tagantsev were able to discriminate between genuine light-induced viscosity changes from those arising due to temperature nonuniformity as a cause of illumination-induced heating.

A great deal of information was provided by Nemilov and Tagantsev concerning the dependence of the photoviscous effect on various parameters such as: (i) the intensity of the incident light, (ii) the wavelength of the radiation which spanned a wide wavelength range (300 - 1200 nm), including sub- and above-bandgap conditions, and (iii) the glass transition temperature, which was reported to decrease considerably upon illumination. The viscosity of the illuminated volume was found to depend exponentially on light intensity, following the form, $\eta(I^{ill}) = \eta(0)\exp(-aI^{ill})$ where *a* is a material dependent constant. A verification of the athermal nature of the effect was the finding that the influence which the light exerts on viscosity decreases at elevated temperature, signifying a counterintuitive or anomalous temperature dependence. The maximum of the photoviscous effect was found to lie in the energy range of localized states. In summary, in the theory of the photoviscous effect, the photo-assisted decrease of viscosity was associated with a decrease in the potential which should be overcome to permit configurational reorganizations. Investigating the photoviscous effect *in situ* in glassy Se by the penetration method, Repka *et al*. [58] reached similar conclusions, as those presented by Nemilov and Tagantsev.

### *12.6.3   Light effects on the glass transition temperature ($T_g$) and the structural relaxation*

Given the context of Section 5.3 and the fact that $T_g$ is typically defined as the temperature where the viscosity attains a value of $10^{13} - 10^{14}$ Poise, it is conceivable that the illuminated volume - becoming more fluid than the equilibrium glass - can be assigned to a decreased "glass transition temperature", $T_g^{ill}$. Stephens was among the first who reported a change (increase) of $T_g$ after illumination of evaporated a-Se films [59]. However, this work cannot be considered relevant to photoviscous effect as the light was used to anneal the film structure prior to the thermal studies. Similar light-enhanced annealing effects of a-Se films were studied by Koseki *et al.* [60,61] whose results showed that the thermal history of a-Se determines the positive or negative change of $T_g$ measured after illumination. Nemilov and Tagantsev found that the difference in glass transition temperature in dark and upon illumination, $T_g - T_g^{ill}$ can reach 25 °C for the AsSe glass [48]. Kolomiets *et al.* [62] examined the photoinduced changes in glass transition in evaporated $As_2S_3$ and As-Se films using an indirect method, i.e. determining the $T_g$ as the temperature at which an initially inscribed scratch on the film surface disappears. It was found that the $T_g$ of the illuminated films increases by 12-15 °C only for the $As_3Se_2$ film, while no photoinduced change on $T_g$ was observed for other compositions. The effect was fully reversible for several cycles of illumination and annealing. The sensitivity of the As-As bonds to light was considered as the main structural origin of the effect.

In parallel, Larmagnac *et al.* [63,64] reported comprehensive investigations of the photosensitivity of structural relaxations of amorphous Se films at $T < T_g$. Data on enthalpy relaxation were fitted by adopting a thermal and a photostructural relaxation time, with activation energies of 372 and 278 kJ mol$^{-1}$, respectively. This result demonstrated a drastic facilitation of the structural relaxation upon irradiation. They also found that the most prominent increase of the relaxation rate takes place for strongly absorbing light (404 nm); an energy which is considerably above the optical bandgap of Se. In contrast, Nemilov and Tagantsev, found the largest photoviscous effect at bandgap light, denoting the absence of the effect for light energy in the deep band transitions. However, in both cases - bandgap and above-bandgap photoelectronic effects on $T_g$ – the penetration depth is quite low to justify volume effects. Notably, both groups offered similar explanation, invoking that photoinduced structural changes diffuse from the processed surface to the bulk volume.

### 12.6.4 Photoinduced effects on elastic constants

Only a few, yet inconclusive, studies have dealt with the athermal role of light on the elastic constants of non-crystalline chalcogenides, studied mainly by monitoring the changes on the sound velocities. Photoinduced changes, manifested as increase in elastic constants, were observed for evaporated $As_2S_3$ films upon bandgap illumination, as demonstrated by the systematic increase in ultrasound velocities of acoustic surface waves [65]. The changes saturate after ~10 min of illumination, i.e. a time scale comparable to that of photodarkening in the same sample. However, in contrast to photodarkening, the hardening of the elastic constants was found to be irreversible. The changes are minor (~6% in sound velocity) and primarily relate to irreversible structural changes of the evaporated film, such as photopolymerization of molecular species. Koman and Klish measured the changes of the (acoustic) speed of hypersonic waves for deposited $As_2Se_3$ thin films near-bandgap illumination (He-Ne laser, $1.25 \times 10^2$ W cm$^{-2}$) [66]. Laser irradiation caused increase in the sound velocity as a result of the elastic constant hardening; the effect was seen to saturate in ~30 min. The $C_{11}$ and $C_{44}$ elastic constants exhibited increase upon illumination by 26 % and 14%, respectively. Trapping of photoexcited carriers at local defect centers was invoked to account for this observation.

More systematic studies were conducted by Boolchand and co-workers who measured the elastic constants in a number of binary $Ge_xSe_{1-x}$ (15 < x < 33) glasses via near-bandgap Brillouin scattering [67]. A significant athermal light-induced softening of the longitudinal elastic constant over a narrow compositional range close to the mean-field rigidity transition composition. The effect is maximized, reaching almost to 50% reduction of the magnitude of elastic constants, at the composition x = 0.19, which is the limit between the floppy and intermediate phase, as shown in Fig. 11. In contrast to the previous-mentioned irreversible photohardening of $As_2S_3$ films detected by an ultrasonic study, the experiment at hypersoninc frequencies for Ge-Se binary glasses revealed a reversible photosoftening phenomenon.

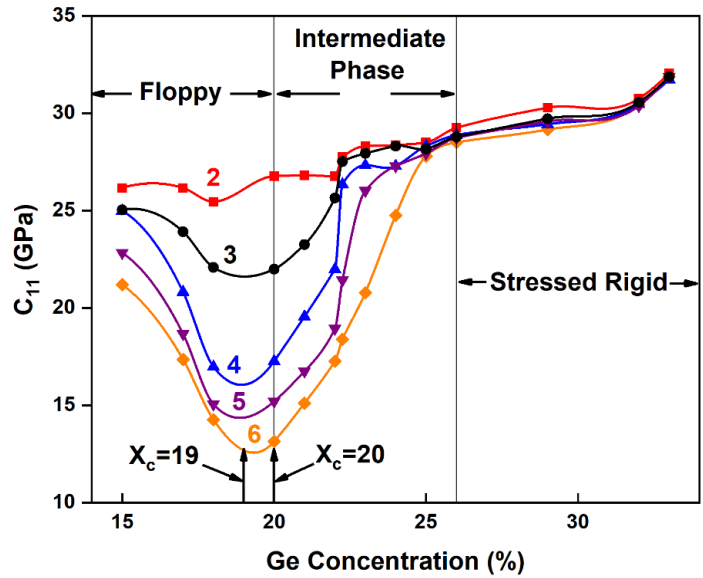

**Fig. 11:** Dependence of the longitudinal elastic constant $C_{11}(x)$ in binary $Ge_xSe_{1-x}$ bulk glasses at various illumination power levels (indicated in mW for each curve). $x_c$ and $x_t$ denote the observed thresholds for photosoftening of $C_{11}$ and the mean-field rigidity transition, respectively. Reprinted from Ref. 67 (figure 4) with permission from [Gump, I. Finkler, H. Xia, R. Sooryakumar, W. J. Bresser and P. Boolchand, Phys. Rev. Lett. 92, 245501 (2004)] by the American Physical Society.

## 12.7 Photoplasticity: combining the photoviscous effect with mechanical stimuli

### *12.7.1 Mechanical and rheological studies*

The photoviscous effect forms the ground for understanding a number of morphic-type photoinduced effects, and in particular, effects that become evident under the simultaneous action of light and an external mechanical stimulus, which facilitate the change of the shape of the illuminated object. A number of *in situ* and *ex situ* studies will be surveyed in this section. Monitoring microhardness changes upon and/or after illumination has been employed as a means to study photoinduced effects into mechanical properties. Asahara and Izumitani found that the illuminated regions of $As_3Se_2$ films become less pliable to scratching than unilluminated areas, providing evidence for hardening of the post-illuminated material [68]. The effect was found to be reversible upon annealing at $T_g$. The photodecomposition mechanism proposed by Berkes *et al.* [69] was put forward by the authors to account for the effect.

Deryagin *et al.* [70] were the first who conducted *in situ* studies to measure microhardness changes upon illumination on thick films of a-Se and bulk glassy $As_2Se_3$, reporting a significant photoinduced decrease of microhardness of about ~20-25% for a-Se and ~50% for glassy $As_2Se_3$. These authors coined the term *photomechanical effect* being influenced by similar studies in crystals [40]. No evidence for a spectral dependence of the magnitude of the photoinduced changes in microhardness was found, while intramolecular bond changes were considered to be at the structural origin of the effect. Kolomiets *et al.* [71] provided quantitative studies, using the Vickers method, on the photohardening effect for $As_{60}Se_{40}$ films. They showed that microhardness oscillates reversibly between two well-defined values upon repeated cycles of illumination and annealing. The role of composition, and hence structure, on the photoinduced hardness changes in $As_xS_{100-x}$ (15≤x≤50) and $As_ySe_{100-y}$ (40≤y≤80) evaporated films was explored by Manika and Teteris [72]. They showed that the modifications in microhardness of films exhibits maxima near x=45 and y=55. The origin of this observation was attributed to photoinduced polymerization of As-Se molecular units (at y=55). Anomalous temperature dependence of the microhardness of irradiated films was observed over a temperature range from ambient up to $T_g$.

The "modern" era of investigations of photoplastic changes using *in situ* microindentation techniques began by Trunov and coworkers in early '90s. A brief review on the subject was published a few years later that described the relevant effects as following a light-driven brittle-to-ductile transition [73]. An important piece of information can be gained by *in situ* studies of photoplastic effects employing micro- and nano-indentation experiments. Photoinduced structural relaxation in evaporated As-S and As-Se films using microindentation kinetic experiments reveal a purely optical photoinduced decrease of the viscosity down to $10^{12}$-$10^{13}$ Poise, which corresponds to the viscosity of the material near the $T_g$ [74–76]. Microindentation techniques exhibit certain drawbacks which do not allow for a complete study of photoinduced elastic-to-plastic transformations, such as the relatively large indented area and the lack of the possibility to determine other important parameters. Depth-sensing instrumental indentation (nanoindentation) was used by Trunov and co-workers to study the photoplastic response in $As_{50}Se_{50}$ evaporated films [77]. *In situ* nanoindentation experiments upon illumination, conducted for as-deposited and annealed $As_2S_3$ films, showed that the penetration depth of the indenter in the former is much deeper (by a factor of 2) than the corresponding one in the latter [78].

Tanaka and coworkers demonstrated quantitative aspects of photoinduced ductility (elongation) in $As_2S_3$ fibers and photoinduced deformation (bending) in $As_2S_3$ flakes by sub-bandgap light under mechanical forces [79, 80]. The term photofluidity was coined to describe the observed morphic effect, which seems improper for this case, given the context of the discussion on photoviscous effect presented in Sections 5.3 and 5.4. Indeed, in conformity to the photoviscous effect and some photohardening experiments, these photoplastic effects exhibited by flakes follow anomalous temperature dependence, lending support to a common microscopic origin of these effects. Notably, qualitative and quantitative different behavior of the temperature dependence of between flakes and fibers was reported [80], as the temperature dependence of the photoplastic effect in fibers is not anomalous, contrary to the results concerning the flakes. The structural differences of $As_2S_3$ owing to the different preparation conditions between flakes (evaporation) and fibers (melt quenching) play presumably a dominant important role concerning the contrasting behavior of the temperature dependence of the photoplastic effect. More recent studies on the mechanical and rheological properties of the material in the course of the photoplastic effect has been conducted by considering the viscoelastic behavior of GeSe$_9$ supporting the transient nature of the effect [81].

### *12.7.2 Raman spectroscopic studies*

A systematic investigation of the photoductility effect in bulk glasses with particular emphasis on the microscopic origin of the effect commenced in our laboratory, almost 18 years ago, after the suggestion of the late Prof. H. Fritzsche who had that time proposed [82] a structural mechanism to account for the finding reported in Ref. [79]. Various binary As-S glasses were prepared in fiber form by drawing the viscous supercooled liquid and were studied by Raman scattering [83–87]. An early review summarizing this work can be found elsewhere [88]. Polarized Raman scattering was employed as a suitable tool to enlighten the structural mechanism underlying photoplastic effects by tracking both the creation and annihilation of certain species and monitor at the same time changes in the orientation of structural units at the short and intermediate length scales. A selection of these results is briefly summarized here.

### *12.7.2.1 Photoinduced orientational ordering*

An important finding emerging from analyzing polarized and depolarized Raman scattering measurements is that orientational ordering progresses in the stoichiometric As$_2$S$_3$ glass as a result of the combined sub-bandgap (632.8 nm) illumination and mechanical (elongation) stress. The glass fibers retain the bulk glass structure whose atomic arrangement, vibrational modes and polarization properties are accurately known, thus facilitating the analysis of the photo-processed material in terms of changes of the above properties.

Representative polarized (VV) and depolarized (HV) Raman spectra recorded from an As$_2$S$_3$ fiber subjected to uniaxial stress are shown in Fig. 12 (left) using a power density of ~20 W cm$^{-2}$. The depolarized spectra exhibit monotonic increase and the depolarization ratio $\rho = I^{HV}/I^{VV}$ saturates at sufficiently large values of the applied stress, $S$. As Fig. 12 (right) shows, this particular dependence of $\rho$ on $S$ was observed for several fibers of various diameters. Data from non-stoichiometric sulfur-rich glasses, which will be discussed below, are also shown. The analysis of a large body of experimental data demonstrated that the measured effect is well reproduced justifying the use of the depolarization ratio $\rho$ as a reliable quantitative indicator of the structural changes that take place in the course of the illumination/stretching procedure. These results are obtained at ambient temperature using illumination sources including sub-bandgap and near-bandgap conditions. *No change in ρ was observed in the absence of the applied stress and vice versa, no fiber elongation takes place under stress, in dark.* The structural changes leading to enhanced orientation remain unchanged after ceasing the illumination/stress stimuli, thus being permanent. Particularly, the form of the curves in Fig. 12 is history-dependent. A sudden stress-jump step from the unprocessed state to a high value of the stress leads to a moderate change in the glass structure orientation. This finding implies that the final state is obtained through a hierarchy of structural mechanisms involved that become activated during the gradual application of the uniaxial stress.

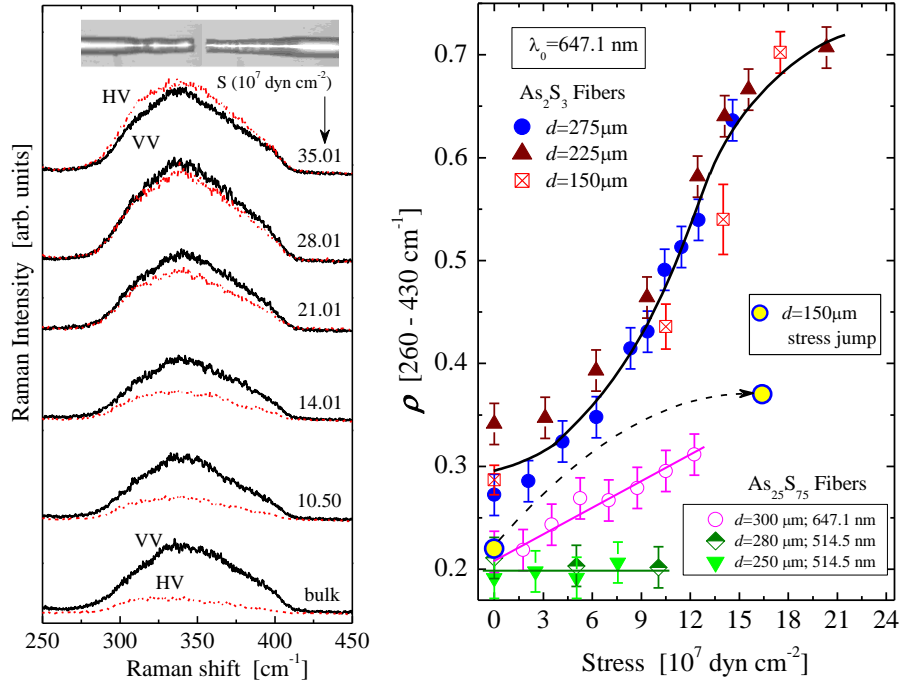

**Fig. 12.** *Left*: Representative polarized (VV) and depolarized (HV) Raman spectra of $As_2S_3$ at various magnitudes of the applied elongation stress. The inset illustrates the moprhic effect experienced by an optically processed fiber under elongation stress. *Right*: Stress dependence of the depolarization ratio, $I^{HV}/I^{VV}$ for $As_2S_3$ and $As_{25}S_{75}$ fibers of various diameters.

### 12.7.2.2 Anomalous temperature dependence – analogy with the photoviscous effect

It has been stressed in previous sections that the photoinduced effects sharing a common origin with the photoviscous effect exhibit striking similarities in the temperature dependence of some property of the effect. In all such cases, the effect under study showed counter-intuitive behavior with increasing temperature, i.e. the effect became less noticeable at elevated temperature.

Performing the experiment described in the previous section at temperatures higher than ambient, albeit appreciably lower than the $T_g$ to avoid heat-induced effects, it was revealed that the photoinduced structural changes responsible for the photoplastic behavior into question, do also follow unexpected temperature dependence. The non-monotonic saturation value of the depolarization ratio is shown in Fig. 13 (left). This saturation value becomes systematically lower for 40 and 60 ºC while for higher temperature it reverts back to the room temperature value (at 90 ºC) and goes beyond that at 120 ºC, thus signifying anomalous temperature dependence. It should be stressed here that the maximum temperature of this experiment (120 °C) is far below the $T_g \approx 210$ °C of $As_2S_3$; hence, precluding thermal induced changes of viscosity. The athermal nature of the effect was further confirmed by heating the fiber under stress in dark and measuring the depolarization ratio at ambient temperature. Neither bandgap narrowing at elevated temperatures is apt to induce heating-induced absorption of the laser light as the decrease of bandgap at 120 °C is of about 3%. The overall behavior of the saturation value of $\rho$, which is a good indicator of the illumination-induced effect of viscosity, is shown in Fig. 13 (*right, up*). The values of $\rho$ correspond to a constant-stress values of $10 \times 10^7$ dyn cm$^{-2}$. The inverse of $\rho$ is a measure of the rigidity of the glass structure or resistance to photoplastic changes. It is noteworthy that this anomalous temperature dependence which is indirectly inferred from spectroscopic measurements shares a striking similarity with the corresponding dependence of the viscosity measured upon illumination, as shown in Fig. 13 (*right, down*). Both curves exhibit maxima at about 60 °C. A detailed explanation considering two competing factors with opposing temperature dependence, leading to the trend shown in Fig. 12, has been presented elsewhere [86].

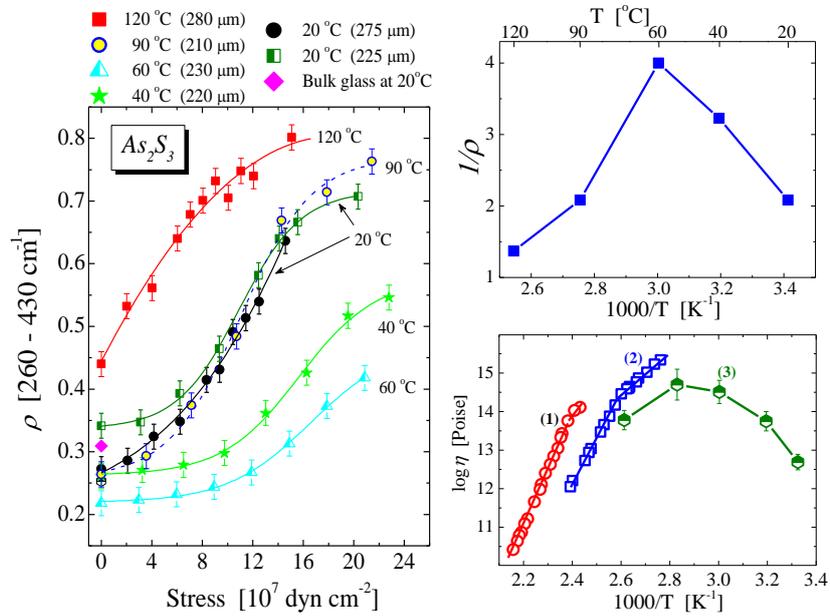

**Fig. 13.** *Left*: Stress dependence of the depolarization ratio for $As_2S_3$ glassy fibers at various temperatures. Dashed and solid lines are guides to the eye. *Right* (up): Temperature dependence of the inverse saturation value of $\rho$ measured at a certain stress value. The data illustrate the anomalous characteristics of the photoviscous effect through the reversal of the temperature dependence above 60 ºC [86]. *Right* (down): Temperature dependence of viscosity of $As_2S_3$; (1) in dark, (2) upon illumination, (3) upon illumination using microindentation measurements. Data for curves (1) and (2) are from Ref. 35; data for curve (3) are from Ref. 73.

### 12.7.2.3  *Photoplastic effects in Sulfur-rich glasses: $As_{25}S_{75}$*

The large body of experimental work on non-crystalline chalcogenides has shown that photostructural changes depend upon the glass stoichiometry, especially in families of glasses where the material exhibits nanoscale-sized *structural heterogeneities*, such as the binary $As_xS_{100-x}$ system [89, 90]. These heterogeneities depend on glass composition and hence different compositions behave in an individual way upon illumination. The structure of the stoichiometric composition $As_2S_3$ is rigid, as it is composed of remnants of layers found in the crystalline counterpart. It also exhibits high chemical order and very limited heterogeneity at the nanoscale. The $As_{25}S_{75}$ glass was explored to examine how photoplastic effects are expressed in glasses with flexible structures containing not only $AsS_{3/2}$ pyramidal units but also the chainlike and ringlike fragments characteristic of sulfur-rich compositions. The selected glass bears the advantage of exhibiting a rich variety of local bonding environments and a relatively high $T_g \approx 140$ °C, i.e. much above the room temperature. The response of this glass to photoplastic changes, using sub-bandgap (647.1 nm) and near-bandgap (514.5 nm) sources ($E_g$ = 2.55 eV), is shown in Fig. 12 (*right*) in comparison to the stoichiometric glass data. The sub-bandgap light induced moderate orientation changes while the near-bandgap one failed to induce any effect. The contrasting behavior between the $As_2S_3$ and $As_{25}S_{75}$ glasses has been discussed by addressing the role of three possible aspects: (i) differences in structure, (ii) the relation between incident light energy and band-gap energy, and (iii) the role of the glass transition temperature [87, 88].

*12.7.2.4 Atomistic mechanisms underlying photoplastic changes in As-S glasses*

Various attempts have been provided to comprehend the atomistic origin of photoplastic phenomena in non-crystalline chalcogenides. However, in their vast majority such models invoke the ubiquitous STE mechanism without providing specific details to the short and medium range structural order of the material under study. Besides, and perhaps more important, such models do not pay attention to the simultaneous emergence of other photostructural changes that modify the structure and interfere with the mobility of the species responsible to photoplastic effect. Indeed, the role of atomic structure is among the most crucial factors, as it has been reported that near-bandgap light causes in sulfur-rich As-S glasses scission of $S_8$ rings and polymerization of the formed diradicals to $S_n$ chains [89, 90]. Such atomic-scale changes at the short and medium range structural order (from zero-dimensional units to arrangements of higher dimensionality) lead to a stiffer glass structure which is less amenable to morphic changes.

The contrasting behavior observed between the $As_2S_3$ ($As_{40}S_{60}$) and the non-stoichiometric $As_{25}S_{75}$ glass call for a deeper elucidation of the role of local atomic arrangement. Sulfur-rich glasses contain chain-like units which are expected to impede the photoplastic effect by reducing the facility of the structure to respond to an external mechanical stimulus. Two such possible scenarios can be envisaged are presented in the sketches of Fig. 14(a) and (b). In the former, possible entangled chainlike sulfur configurations will hindering the flow process. In the latter, such chainlike fragments can be reconfigured (relaxed), upon illumination and stressing, to directions other than that of the applied stress; hence not contributing to the fiber morphic changes. Evidently, structures with 3-dimensional network structures will be less responsive to photoplastic changes owing to the over-constrained atomic arrangement and the lack of *weak* bonds, which are primarily susceptible to illumination. As a result, the specific structural characteristics and the concomitant softness of locally layered materials, such as $As_2S_3$, enhance the feasibility of structural changes that lead to the plastic deformation, Fig. 14(c).

Fritzsche was among the first who proposed that *intramolecular* bond rupture can account for such effects [82]. He proposed that such atomic changes leading to photoplastic effects are common to any chalcogenide at any temperature, which was shown (see above discussion) that it is not entirely correct. Based on the spectroscopic data, an intramolecular model describing the opening and incorporation of $As_4S_4$ cagelike species into the glass structure was proposed [87].

This model can partly offers a quantitative estimation of the fiber elongation upon illumination and stress, which amounts up to ~30% of the observed effect. A schematic of this structural model is depicted in Fig. 14 (e-g). The consideration of the specific role of the As-As bonds in the photoinduced transformation of the $As_4S_4$ cagelike species into orpiment-type configuration revealed that there exist in the glass *sub-structures* (of nanoscale extent) which could be "in resonance" with sub-bandgap light [87]. In accord to this reasoning, recent molecular orbital simulations on Se have shown sub-structures (curled and intersecting Se chains) with lower "bandgap" than the bulk can be easily affected by sub-bandgap light and thus have been envisaged as possible candidates facilitating photoplastic effects [91].

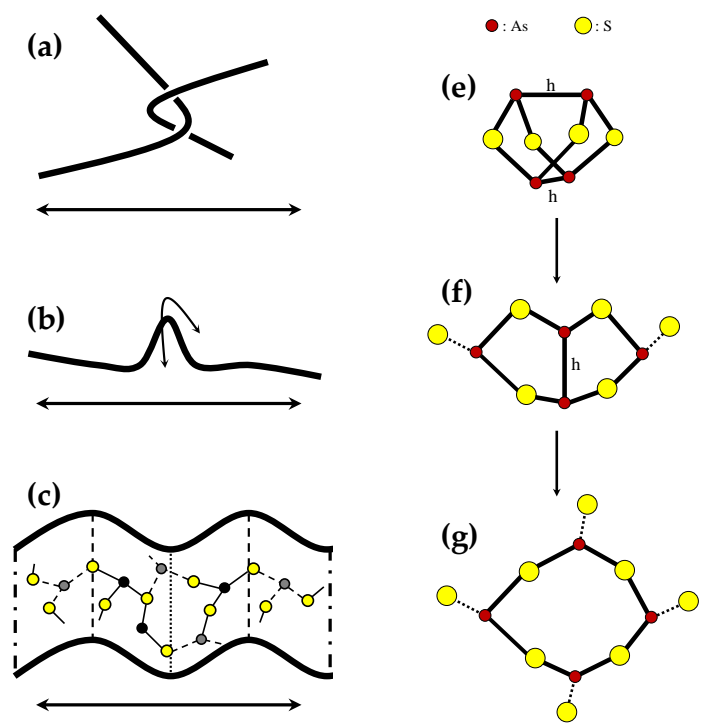

**Fig. 14.** *Left*: Schematic representation of possible local structural arrangements in non-stoichiometric S-rich structurally flexible glasses **(a)** and **(b)** and in the $As_2S_3$ glass **(c)**. The double-side arrow portrays the direction of a hypothetical external elongation stress. *Right*: A possible way of a locally-planar morphogenesis. Highly symmetric realgar-type molecules ($As_4S_4$) unfold, after the rupture of a homopolar (h) As-As bond, giving rise to planar-like configurations. Reprinted from Ref. 88. Copyright © 2003, John Wiley and Sons.

Last but not least, the proliferation of the structural changes from the short-range to the medium-range inevitably necessitates the adoption of *intermolecular* structural models to account for photoplastic effects. The buckling model originally developed by Ihm [92], has frequently been considered to describe structural changes in chalcogenides. The model is grounded on the layered structure of chalcogenides and the weak inter-layer interactions i.e. van der Waals bonds, which are easily responsive to light.

### *12.7.3 Photoinduced mass transport: relief gratings and healing effects*

The first who reported and accounted for photoinduced effects related to mass transport, studying the kinetics of "healing" processes of scratches inscribed on the surface of chalcogenide films, was Kolomiets *et al*. [93]. The observed effect was manifested as an oscillating, reversible change - during illumination and annealing cycles - of the temperature needed to heal a channel scratched on a chalcogenide film. The optically field-induced mass transport resulting in giant relief modulations formed on the surface of a-$As_2S_3$ upon intense ($5\times10^2$ W cm$^{-2}$) bandgap light illumination, was attributed to the photoinduced softening of the glass matrix leading to the formation of defects with enhanced polarizability [94]. To account for this effect, it was suggested that these defects experience diffusion under the optical field gradient force generating surface relief modulations. The effect is realized either by single- or two cross-polarized beam experiments and depends upon polarization of the electric field. Poborschii *et al*. [95] reported the emergence of relief deformations for a-Se under bandgap illumination at low temperature. This observation was assigned to photomelting taking place via an interchain bond breaking mechanism.

Trunov and co-workers have systematically explored similar effects for a number of non-crystalline chalcogenides [96–100]. They demonstrated that similar mass transport effects can be achieved even at considerably lower power densities. This is a useful observation because such reduced light fluence is in the range of light intensities employed for the recording of holographic gratings in non-crystalline chalcogenides. Various details of the photoinduced mass transport were studied both experimentally and theoretically (to obtain photoinduced diffusion coefficients), including the inversion of the direction of the surface relief grating, kinetics of the effect at low temperature, and selective light-induced mass transport in amorphous $As_xSe_{100-x}$ tuned by composition see Ref. [101] and references therein. Some recent attempts are also focused in the formation of nanoscale structures in amorphous chalcogenide films [102]. This can be achieved by controlling appropriately the conditions by incorporating noble metal nanoparticles into the chalcogenide structure and using the excitation of the plasmon resonance of the nanoparticles. Finally, the expression of surface relief gratings using electron beams has also recently been emerged as an interesting phenomenon for manipulation the texture of non-crystalline chalcogenides at the nanoscale [103, 104].

## 12.8   Athermal photoinduced changes of the phase state

Photoplastic or morphic effects, described in the previous section, are perhaps the most fascinating photoinduced phenomena as they emerge with spectacular changes in the shape of the material. However, photoinduced changes of the phase state can be even more intriguing as they seem to violate in certain cases the common wisdom. Their *athermal* nature and astonishing reversibility in some circumstances, entails prospects for captivating applications. Such effects are perhaps the most challenging to comprehend and explain from the wealth of photoinduced phenomena in non-crystalline chalcogenides.

The current section is divided into two sub-sections. In the first, a brief overview of amorphous-to-amorphous transitions - the subject of which is quite limited - will be reviewed, while in the latter we will survey on the more extended literature on reversible amorphous-to-crystal photoinduced transitions.

### *12.8.1   Reversible amorphous-to-amorphous transitions (AAT)*

In analogy to the concept of polymorphism observed in crystalline solids upon changing the temperature (T) or the application of pressure (P), non-crystalline materials can also exhibit transitions to a distinct amorphous phase. While in crystals, both T and P are parameters that are frequently used to drive and tune the phase transition, in amorphous solids, it is predominantly pressure that can drastically change the short (nearest neighbors) and medium (e.g. number of ring members) range structural order, leading to a new amorphous phase. Mishima *et al*. [105] was the first who pointed out that an amorphous-amorphous transitions exists in water. Several types of disordered solids experience such pressure-induced transitions, including also chalcogenide glasses, such as $As_2S_3$ [106] and Ge-Se [107]. A "hidden" thermally-driven phase transition within the amorphous state of $As_{50}Se_{50}$ was suggested to account for the findings of calorimetric studies, where two exothermic peaks, situated below and above $T_g$ were observed [108]. The idea was based in the view that the glass and the parent crystal have similar short-range order. It was thus suggested that two different glassy structures can (co)exist in the $As_{50}Se_{50}$ composition, each one related to the low- and high-temperature crystalline polymorphs. Under this situation, illumination could possibly induce transitions from one structural configuration state to the other [108].

Reports on well-documented optically-driven AATs are very rare [109]. Typically, an ATT cannot be considered as an actual phase change transition for two main reasons. First, an AAT is primarily related to subtle differences in the atomic structure of the two distinct meta-stable structures in the amorphous state. Second, the two distinct phases are not determined as in crystals by thermodynamics; they are both non-equilibrium states and the details of the stimulus (e.g. photons) employed to transform one phase to another strongly affects the kinetics of the transformation. It is worth-noting that whereas the atomic change can be moderate, an AAT can give rise to substantial changes in the optical properties, i.e. enhanced optical contrast between the two amorphous phases. This effect can endow new functionalities, possibly exploitable in next generation optical data recording, photonic bandgap tuning, and so on.

Given that structural changes accompany practically any photoinduced effect, then a plausible question arises: what is the fraction of the atoms/bonds that must be affected in order to consider this structural change as an AAT? Typical photoinduced effects such as photodarkening or photobleaching have been considered to involve a limited fraction of the atoms in the illuminated volume, which does not exceeds ~2% [26]. This small fraction does not justify the term AAT in this case. Notably, there are few cases where As-S binary chalcogenide glasses exhibit massive photo-induced bond reconfigurations. Detailed spectroscopic studies have shown that such Sulphur-rich binary glasses the fraction of bonds altered by illumination can be astoundingly high, ~20%, that is, almost one order of magnitude higher than the corresponding photo-induced reorganized bond fractions reported up to now in typical cases [89, 90]. The main structural transformation in these glasses involves the photoinduced scission of $S_8$ rings and the polymerization of these diradicals to chain-like species. However, photoinduced polymerization is practically a transient effect and hence not classified as an AAT. Nanoscale phase separation was considered to be the key for this extraordinary behavior leading to massive photoinduced structural changes. Nanoscale phase separation can be envisaged as the situation where the composition of the particular system under study dictates a structure where local micro-environments of particular atomic configurations exist with comparable internal energies. Therefore, external stimuli can rather easily induce transformations from one local structure to another.

The exploration of non-crystalline chalcogenides with high responsiveness in illumination and hence potential to applications, is a guide towards considering materials compositions which exhibit structural motifs that are able to withstand and proliferate the photoinduced changes into their volume, in a sustainable aspect. Evidently, this can be achieved in cases where the disorder characterizing the structure can afford certain relaxation routes which can lead to relaxation of the photoexcited bond to configurations different than the initial one. Given this context, disordered materials exhibiting nanoscale phase separation, as described above, are possible candidates for exhibiting structural transformations among two structural motifs, with alike bonding features, at low-cost of energy.

Arsenic-rich, binary As-Se amorphous films grown by the pulsed-laser deposition technique were shown, using x-ray photoelectron spectroscopy (XPS), to exhibit strong nanoscale phase separation [110, 111]. This particular deposition method plays significant in the growth of amorphous chalcogenides with tailored-made functionalities owing to its advantage of maintaining quite well the composition of the target material, avoiding also the formation of structures and morphologies associated with irreversible processes. In brief, XPS data were interpreted in the context that the structure of $As_xSe_{100-x}$ PLD films ($x > 40$ at.%) consists primarily of $AsAs_{3-m}Se_m$ pyramidal units. The units $As_4$ ($m = 0$) and $As_3Se$ ($m = 1$) are denoted as type-$As^I$ environment. The As atoms in $As_3Se$ pyramids have alike electronic properties with $As_4$ units, hence, termed as *elemental-like* domains. Based on similar grounds, pyramids $As_2Se_2$ ($m = 2$) and $AsSe_3$ ($m = 3$) constitute the *stoichiometric-like* domains; type $As^{II}$ environment. The fraction of As atoms participating in the elemental-like environment ($As^I$) was found to be in range 15−20%.

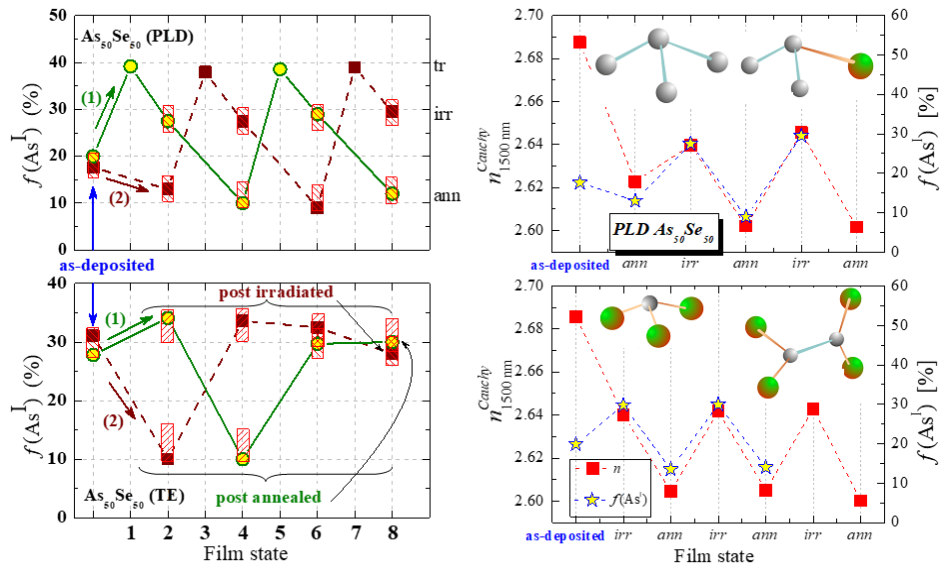

**Fig. 15.** *Left*: Switching of fraction $f(As^I) = As^I/(As^I + As^{II})$ (%) in $As_{50}Se_{50}$ thin films deposited by two different techniques. PLD (up) and TE (down) after imposing alternating external stimuli. The film states 1–8 correspond to cycling of irradiation followed by annealing or *vice versa*, where route (1) starts with irradiation and route (2) with annealing. The three lines of data points (up) shows the states: (i) transient (tr, laser ON), (ii) post irradiated (irr) and (iii) post annealed (ann). *Right*: Correlation between changes in refractive index (*n*) at $\lambda = 1500$ nm (measured by ellipsometry) and $f(As^I)$ for as-deposited $As_{50}Se_{50}$ PLD films after imposing external stimuli in the sequence irradiation → annealing (up) and annealing → irradiation (down). The inset shows a schematic of structural *elemental*-like ($As^I$) and *stoichiometric*-like ($As^{II}$) pyramids. Reprinted from Ref. 91. Copyright © 2012, John Wiley and Sons.

Exploring in detail structural and optical properties of $As_{50}Se_{50}$ PLD and TE films it was found (see Fig. 15) that reversible switching takes place between two structurally distinct amorphous states, guided by exposure to near-bandgap light (increasing $As^I$) and annealing (decreasing $As^I$). This behavior justifies the use of the term AAT, especially in view of the large fraction of atoms involved and the concomitant change in the optical properties of the films. The oscillating effect does not depend on which external stimulus (light and annealing) is imposed first. Notably, only in PLD-prepared films the effect is observed repetitively, for several illumination/annealing cycles, while this does not hold for TE films.

Another interesting outcome emerging from this study is that the photoinduced structural changes bear a rather strong transient character, as revealed by the film states "1" and "5" in Fig. 15 (left panel). As long as illumination is "on" the fraction of $f(As^I)$ reaches about 38−40%, while it decreases after ceasing illumination to ∼30%. In addition, annealing always causes considerable reduction in $f(As^I)$ to a value near 10%, originating from the rupture of As-As bonds and the formation of $As_2Se_2$ and $AsSe_3$ pyramidal species.

Figure 16 summarizes in a pictorial way the structural transformation of the AAT described above in the frame of an energy landscape representation of the involved steps. While in a typical reversible amorphous-to-crystal transition (exploited in optical data storage) the atomic arrangement oscillates between the crystal (state "1") and the glassy state (obtained by melting and rapid quenching, state "2"), a richer phenomenology is offered in materials exhibiting nanoscale-phase separation which can drive the structure to amorphous states with alike, albeit discrete, bonding features.

This type of AAT transitions could possibly be a more generic phenomenon in disordered materials, which however awaits to be studied. As discussed, the presence of necessary populations of structural species is a prerequisite enabling an external stimulus, such as light, to covert species athermally from one population to another. The refractive index contrast of $As_{50}Se_{50}$, under the condition of the current experiment amounts to $\Delta n \approx 0.04$, is only a factor of two lower than the contrast in $Ge_2Sb_2Te_5$ at $\lambda = 633$ nm (DVD operating wavelength), implying high potential for applications. Such structural rearrangements would entail low-energy consumption in potential devices as the ruptured and formed bond energies are comparable i.e. 200 kJ mol$^{-1}$ (As-As) and 230 kJ mol$^{-1}$ (As-Se). In comparison to crystal-amorphous transitions exploited in phase-change memories, the ATT entails negligible volume change and avoids films stress during crystallization which may suppress the growth rate. Exploitation of the diversity of distinct structural states in disordered structures could also provide possibilities for multi-lever data storage mechanisms. However, viable applications are still elusive as many details of AAT as processes for data storage are still unknown.

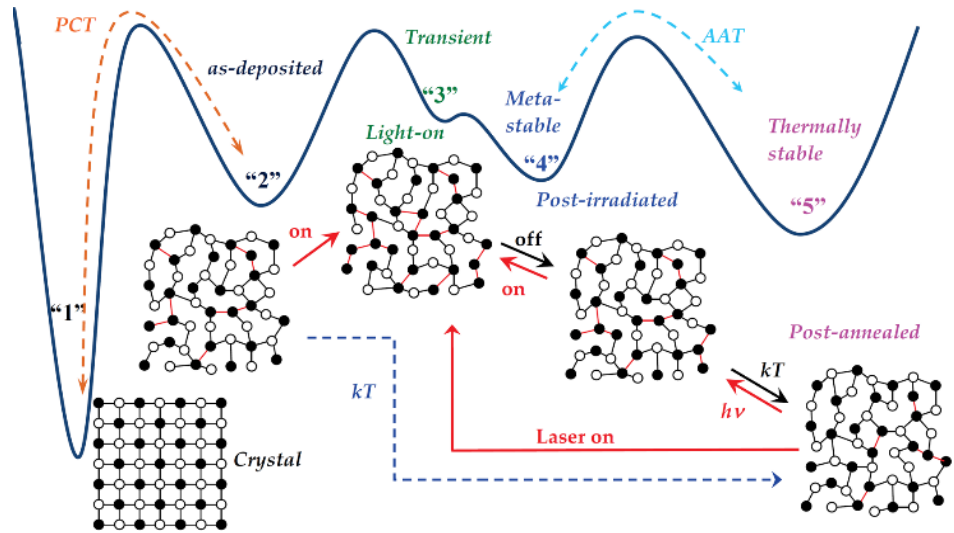

**Fig. 16.** Potential energy landscape representation of the photo- and heat-induced transitions in $As_{50}Se_{50}$ PLD films. The deepest minimum "1" corresponds to the crystal which has been (depicted here as an idealized structure, not reflecting the accurate crystal structure). The amorphous as-deposited state "2" is built-up by both As-Se heteronuclear (black lines) and As-As homonuclear (red lines) bonds. Illumination guides the structure to minimum "3", where the number of As-As bonds increases; the system is transiently trapped in a shallow minimum being prone to relaxation into the post-irradiated state "4", characterized by a density of As-As bonds larger than "2" but lower than that of "3". State "4" is thermally meta-stable; after annealing it relaxes to state "5" which is the amorphous state with the lowest density of homonuclear As-As bonds. Reversible switching in the structure and the optical properties of PLD films occurs between states "4" and "5". State "5" can also be reached from state "2" after annealing. Reprinted from Ref. 91. Copyright © 2012, John Wiley and Sons.

### *12.8.2 Athermal transitions between amorphous and crystalline states*

#### *12.8.2.1 The framework of phase changes by electronic excitation*

From a different perspective, irradiation-induced phase transitions in materials, or more generally stated, modifications of materials due to electronic excitation, is a generic effect observed not only in non-crystalline chalcogenides but, in principle, in any crystal [112]. Irradiation, by photons (laser annealing) and less commonly by electrons, has been used as a versatile means to modify the material's surface to anneal damage created by various processes, such as implantation. For sufficiently intense beams a high density of photoexcited carrier are created, called electron-hole plasma. A key concept in understanding such phenomena is that the temperature of the excited electrons and ions is not necessarily the same [112].

Perhaps the first attempt to explain a non-thermal solid-liquid transition, a kind of *athermal melting*, traces back to the ideas developed by Van Vechten and coworkers based on the photoexcited electron-hole plasma [113, 114]. In an effort to account for critical observations, which had gone unnoticed in previous studies of the pulsed-laser annealing of crystalline Si, concepts for non-thermal transformation of the material's surface from crystal to liquid were developed. In brief, it was postulated that if the density of photoexcited electrons, which is materials dependent, exceeds a certain threshold, then a second-order phase transition will take place. The depletion of bond charges causes instability of transverse acoustical phonon modes to the extent that the crystal cannot sustain shearing, hence passing to the "fluid" regime. In contrast to the normal thermal melting where the crystal undergoes a first-order transition and the atoms carry out long oscillations around the equilibrium positions, the athermally induced fluid retains its energy in the electrons, maintaining the atoms rather restful. The duration of the laser pulse is critical. For short durations, i.e. comparable or less than the vibrational period the effect in purely athermal as thermalization has not been settled.

#### *12.8.2.2 Early studies on photocrystallization*

As thermal effects are practically unavoidable during illumination of a chalcogenide material, a great deal of confusion surrounds studies of photocrystallization. Literally speaking, photocrystallization arises solely from photoelectronic processes. Essentially, the sharp increase in the free carrier concentration upon illumination and the excitation of electrons across the mobility gap gives rise in a large density of broken bonds in the material. This process weakens the metastability of the amorphous state rendering the structure more amenable to crystallization.

Dresner and Stringfellow were the first who studied the role of illumination in the characteristics of the phase change in elemental glassy Se [115]. They made reference to previous similar studies where illumination-enhanced crystallization rate of amorphous Se was interpreted based on thermal or combined photo- and thermal effects. Measuring the growth rate of the diameter of individual crystallites (spherulites) upon irradiation with a 100 W Hg lamp, they were able to obtain the crystallization kinetics. It was suggested that the effect of illumination on the rate of crystal growth is related to the generation of hole-electron pairs. Holes are the deeply trapped carriers that dictate the diffusion of the photoexcited pair, thus the flux of free holes are considered to control the growth rate of the crystal. The main contribution to the athermal mechanism of photocrystallization was ascribed to the energy released upon annihilation of the hole-electron pair, which could provide a means for reorientation of the Se chains disposing them in a crystalline configuration. Several other studies were focused on photocrystallization of various non-crystalline chalcogenides, i.e. Se [116], $Ge_x$-$Se_{1-x}$ [117], $Se_{85}Te_{15}$ [118], and $Se_{100-x}Te_x$ [119].

Ovshinsky and co-workers [120, 121] showed that msec long light pulses induce crystallization and re-amorphization of $Ge_xTe_{1-x}$ (x=0.11, 0.66 and 0.72); however, they did not attempt to separate optical and thermal effects. Feinleib *et al*. [122] studied reversible light-induced crystallization of the amorphous $Te_{81}Ge_{15}Sb_2S_2$ compound. This effect was considered as the optical analogue of the memory-type electrical switching discovered earlier in such compositions by S. R. Ovshinsky [123]. Feinleib *et al*. [122] suggested a model to interrelate the optical and electrical switching effects, proposing that the optically-induced reversible phase change does not solely originate from thermal effects but is also influenced by the generation of the photoexcited electron-hole carriers. Few more reports on the laser-assisted reversible crystallization of Te-rich alloys stated that the distinction of the role of optical and thermal effects was yet not clear [124, 125]. No progress in this field took place for a decade or more when athermal reversible quasicrystallization in $GeSe_2$ glass studied by Raman scattering was reported by Griffiths *et al.* [126, 127]. The authors used various levels of the laser power interpreting the spectra as showing a transformation of the structure to a quasicrystalline configuration before it is converted to microcrystallites. They considered that photoinduced crystallization of the $GeSe_2$ glass is reversible after keeping the glass in dark for several hours. The ambiguous interpretation of the Ge-Se Raman modes in that work made the elucidation of the mechanism underlying this photoinduced process dubious. In a more systematic experimental study by Haro *et al*. [128] light-induced crystallization of $GeSe_2$ was slightly reconsidered. The "cluster" and "microcrystallites" approach interpretation employed by the authors led them to debatable interpretations of the experimental data, as strong Raman bands were erroneously assigned to Ge-Ge and Se-Se bonds in the stoichiometric glass. The use of above-bandgap light is another weak point which does not preclude the possibility for a significant influence of thermal effects. Indeed, the photocrystallization effect exhibited much slower kinetics with sub-bandgap light owing to the avoidance of heat-induced effects. Significant Raman line broadening is an additional solid evidence for thermal effects.

On another front, quite a few studies were dedicated to elucidate the influence of light polarization on the photocrystallization of a-Se. Irradiation at $T > T_g$ was found to induce anisotropy due to Se chain orientation [129]. Obviously, at such temperature, the effect has a strong thermal character in addition to the optical effect. Lyubin *et al*. [130, 131] studied the effect of laser polarization on the crystallization behavior of Se-containing amorphous chalcogenide films. It is not rare to meet contradictions in photocrystallization studies in the case of a-Se. As this material has its $T_g$ at about room temperature, even small differences in the illumination power density can qualitatively and quantitatively change the features of the photocrystallization process, i.e. kinetics, preferred orientation, structure, and so on. For example, Roy *et al*. [132] explored the effect of illumination wavelength on the photocrystallization of a-Se using above- and sub-bandgap light. They observed a preferential growth of the crystal by sub-bandgap light (676.4 nm) illumination and non-preferential growth by the above-bandgap light (488.0 nm) using 120 W cm$^{-2}$. This observation contradicts the findings of Innami and Adachi [133] using 488 nm illumination at 20 W cm$^{-2}$ who reported preferential orientation for this wavelength.

### 12.8.2.3 *Early studies on photoamorphization*

A large number of articles have been published on the study of photocrystallization in amorphous chalcogenides, as discussed in the previous Section. This effect is more feasibly realized as dictated by thermodynamics, since the crystal is the most energetically favored state. Besides, thermal effects can intervene – in parallel to light-induced effects – especially for low-$T_g$ materials for which the supercooled regime ($T > T_g$), where crystallization rate is maximized, is easy to attain. On the other side, photoamorphization, i.e. the light-assisted disordering of a crystal is a more uncommon effect. When observed, it inherently entails an athermal origin as there is no way of direct transformation of a crystal to an amorphous solid, unless melting (necessitating substantially high T's) intervenes.

Amorphization of a crystal can occur via high pressure, where the pressure quenched product can retain to a large extent the disorder induced under *in situ* pressure conditions [134]. Amorphization of materials, which can either form glasses [135] or not [136], can also take place under irradiation by energetic particles, such as neutrons. However, achieving a crystal-to-amorphous transformation by a "gentler" and more convenient stimulus, such as light, is more intriguing as this structural transition goes beyond our current understanding of disordered systems and deserves thorough consideration.

Few studies have up until now focused on athermal phase change transitions of chalcogenides. Kolobov and Elliott reported that thermally evaporated amorphous films of $As_{50}Se_{50}$, after being thermally crystallized at 140 °C, can be retrieved back in the amorphous form under low-intensity white light illumination for a duration of 150 min [137–141]. This is the sole composition of the binary system As-Se exhibiting the athermal transition. The amorphous-crystal-amorphous cycle can be repeated consecutively. However, only the crystalline-to-amorphous transformation step (photoamorphization) is achieved optically. Partial amorphization was observed even in the case where the film was illuminated at high temperature (140 °C). The kinetics of photoamorphization yielded an activation of 0.15 eV was estimated. Evidence for the *intramolecular* origin of the effect was provided by comparing the Raman spectra of the crystallized and amorphous films revealed that the molecular realgar-type $As_4Se_4$ species found in the crystal, are not present in the amorphous phase, which implies a photo-polymerization reaction leading to a network-like solid. Another possible mechanism of *intermolecular* origin was envisaged according to which the cage-like molecules $As_4Se_4$ are not destroyed, but these units just alter their relative orientation and spatial distribution upon illumination as a result of weakening of intermolecular interactions. This is a process resembling thermally-induced orientational transitions in plastic crystals. Based on the disappearance of the sharp Raman band of $As_4Se_4$ molecules in the amorphized film [139], the intermolecular model seems more plausible.

Photoamorphization of orpiment, $c-As_2S_3$, was reported by Frumar *et al.* [142] Due to the extreme stability of supercooled $As_2S_3$ and its inherent difficulty towards crystallization to the orpiment phase, the authors studied natural crystals, which are not free of contamination. Using near-bandgap light they observed partial only crystallization of the crystal at room temperature. Similar effects were observed at low temperature (36 K), confirming the athermal nature of the effect. Frumar, *et al*. [142] considered that the softness of the layered orpiment crystal can cause the generation of several defects under mechanical treatment. The presence of a high defect density can endow the metastable crystal with a Gibbs free energy *higher* than that of the amorphous state. Hence, light can provide the required energy needed to obtain the transition from the metastable (defected) crystal to the disordered state.

### 12.8.2.4  Recent progress in athermal, optically-driven, reversible amorphous-to-crystal transformations

A recent detailed study of athermal photoinduced transitions between the amorphous and the crystalline states has been conducted by Benekou *et al*. [143] for the multicomponent bulk glass $Ge_{25.0}Ga_{9.5}Sb_{0.5}S_{65.0}$ containing 0.5 at.% of $Er^{3+}$ using above-bandgap laser illumination. Essentially, this the first study reporting an all-optical, truly reversible amorphous-to-crystal transformation where the complete cycle is achieved by illumination, without involving annealing steps. The benefit of the time-resolved Raman measurements (Fig. 17), which was employed, is that the photostructural changes were monitored in detail in the time scale of few seconds. The observed effects were purely athermal, as the temperature rise of the illuminated volume was estimated from Raman band shifts to be far below the $T_g$ of this glass. A reliable assignment of the Raman bands is indispensable for the interpretation of the spectral changes generated in the course of illumination, which has been the main drawback in previous studies. Figure 17 (left) shows the Raman spectrum "1" of the well-equilibrated glass recorded at low light fluence unable to engender photostructural changes. The spectrum is fitted with various lines, each one assigned to particular species that can be found in the structure at reasonable concentrations for the glass composition under study [143]. CST and EST stand respectively for corner-sharing and edge-sharing tetrahedra. Bonding arrangements of such tetrahedra on medium-range order and beyond, develops structures of 3-D and 1-D topological character.

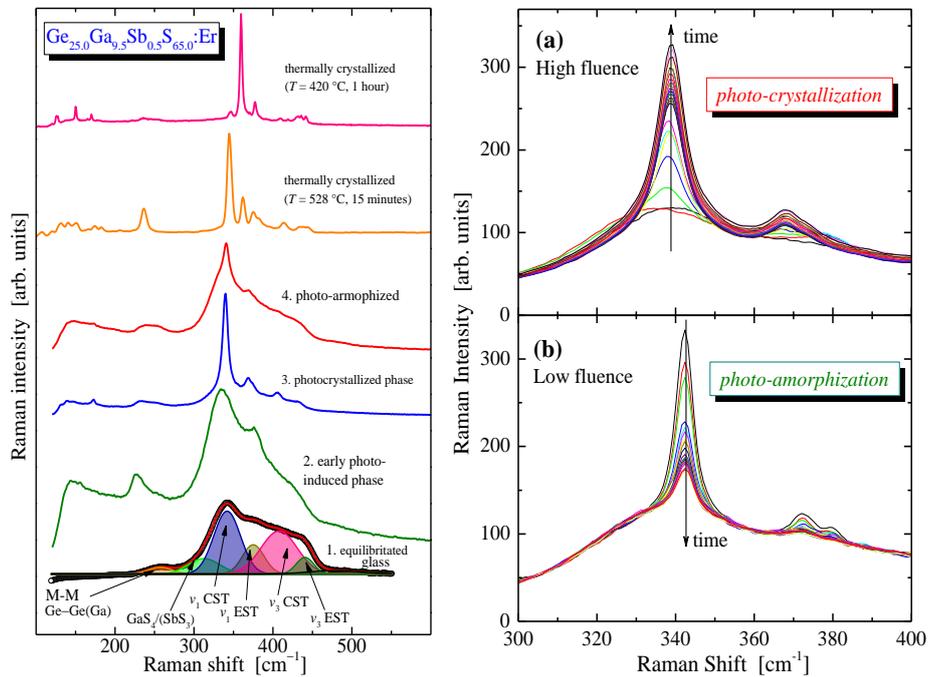

**Fig. 17.** *Left*: Raman spectra showing the various stages of photostructural changes (from "1" to "4") under continuous illumination. The spectra of crystals obtained by annealing above $T_g$ are shown on top. *Right*: Normalized time-resolved Raman spectra showing **(a)** the photoinduced crystallization step dominated by the $A_1$ mode, ~338 cm$^{-1}$, and **(b)** almost complete re-amorphization after ~18 minutes of continuous low-fluence illumination. Reprinted from Ref. 143. Rights managed by AIP Publishing.

The analysis of Raman spectra revealed that a *three-stage mechanism* of photostructural changes takes place at prolonged illumination. During the *first stage* of illumination significant photostructural changes were observed within the glassy state, prior to any photocrystallization effects. Spectrum "2" in Fig. 17 (left) Shows the significant enhancement of two new bands situated at ~235 cm$^{-1}$ and ~375 cm$^{-1}$. The band at 235 cm$^{-1}$ originates form Ge-Ge(Ga) ethane-like configurations. These units are formed as light ruptures CST and/or EST. The increase of the band at 375 cm$^{-1}$ in spectrum "2" makes clear that illumination causes increase in the EST/CST ratio during this early illumination stage.

These structural changes start immediately with illumination and saturate in about 1-3 min for $P$ ~10$^4$ W cm$^{-2}$. This transformation has an important outcome; it entails degradation of the 3-D network to structures of 1-D character (EST and ethane-like units). Hence, more available space or free volume is created, which is essential for triggering and facilitating atomic displacements towards the *second stage*, i.e. crystallization. Indeed, after this "incubation" first period, crystallites start growing as Raman spectra show. Spectrum "3" in Fig. 17 (left) stands for the final photo-crystallized product. The very weak background in the spectra range 300-400 cm$^{-1}$ shows that almost complete photocrystallization has been achieved. The kinetics of photocrystallization is quantified by evaluating the contribution of the Raman band at ~340 cm$^{-1}$, which reflects the $A_1$ of the low-temperature 3-D structure $\beta$-GeS$_2$ crystal located at the same energy, Fig. 17 (right, (a)). Photocrystallization saturates in about 10 min. Changes in the Raman wavenumber (redshift by 1-2.5 cm$^{-1}$) caused by laser heating are translated to temperature increase of about 60-150 K, i.e. much lower than the $T_g$ of the glass under study.

Perhaps, the most intriguing observation is that by lowering the illumination power by one order of magnitude ($P$ ~10$^3$ W cm$^{-2}$) the photo-crystallized volume can be re-amorphized. The intensity of the Raman bands of the crystalline phase weaken progressively until the final stage, shown by spectrum "4" in Fig. 17 (left).

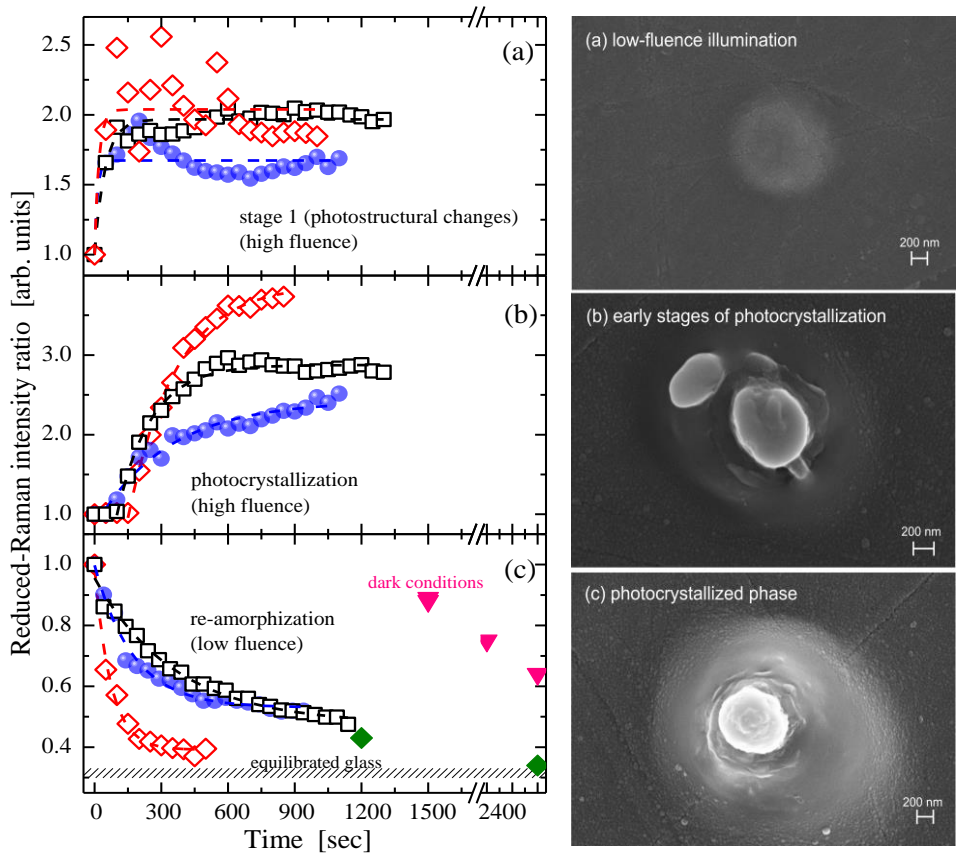

**Fig. 18.** *Left*: Time-dependence of selected normalized reduced Raman bands describing the kinetics of the three stages of photostructural changes: **(a)** stage 1 – photostructural changes; **(b)** stage 2 – photocrystallization; **(c)** stage 3 – re-amorphization (low fluence). The dashed area indicates the value of the equilibrated glass. Three independent measurements from different illuminated spots are shown by the symbols where the corresponding symbols across the panels represent the same spot, and the dashed lines are exponential fits to the data points. Independent measurements of the slow re-amorphization under dark conditions (solid triangles) and almost complete re-amorphization in ~60 minutes under low-fluence illumination (solid diamonds) are also shown in (c). *Right*: Representative SEM images; see text for details. Reprinted from Ref. 143. Rights managed by AIP Publishing.

The kinetics of the photoinduced re-amorphization is studied in detail (Fig. 17, right, (b)), revealing a time scale of ~60 min. The re-amorphized product retains only a weak crystalline component. A quantitative aspect of the kinetics of the three steps involved in the photoinduced cycle glass-crystal-glass is presented in Fig. 17 (left). Details are given in the figure legend. It is worth-mentioning that the photo-crystallized product shows a weak tendency towards re-amorphization even in dark (laser-off). However, the kinetics in this situation is at least two orders of magnitude slower than the corresponding kinetics upon continuous illumination with low-fluence light.

An interesting aspect of the observed fully reversible cycle (glass-crystal-reamorphized solid) is that this cycle can be repeated in a row for more than one time yielding similar results. The photocrystallization step is quantitative different for the various cycles, while the re-amorphizaiton process exhibits practically the same kinetics. In addition, the aforementioned reversible structural changes are accompanied by irreversible morphological changes reminiscent of those observed in the photoinduced mass transport phenomenon. Figure 18 (right) displays scanning electron microscopy (SEM) images during the photocrystallization process. Image (a) shows a faint imprint of the laser beam on the surface of the glass induced at low-fluence conditions which do not cause photostructural changes. Changes in surface morphology of the glass developed at high fluence are depicted in SEM images (b) and (c), which correspond to the early and final stages of photocrystallization, respectively. Mass transport results in the formation of columnar-like structures with dimensions of 1-1.5 μm.

More systematic studies are currently underway in an effort to decipher the role of the Sb/Ga ratio in $Ge_{25}Ga_xSb_{10-x}S_{65}$ glasses ($x$ = 5, 7.5, 10) in the reversible athermal transition between the crystal and the glass [144]. Remarkably, the pure $GeS_2$ glass does not exhibit this athermal photoinduced transformation, which manifests the crucial structural task of Ga/Sb in the glass. The creation of local (nanoscale) environments resulting from the different bonding requirement of the Ga/Sb atoms (3-fold coordination to S) is apparently a decisive factor for the responsiveness of the multicomponent glass to light. Despite quantitative differences, all glass of the above quaternary system demonstrated the athermal, reversible crystal-to-amorphous transition.

## 12.9  Summary and perspectives

In the preceding sections we have tried to delineate the particularities and microscopic origin of a number of photoinduced effects in non-crystalline chalcogenides. Emphasis was placed to categorize and sort effects of common origin. As a general perception, we should stress that the photoviscous effect, i.e. the athermal decrease of glass viscosity under illumination, stands as the cornerstone of a number of photoplastic effects. Macroscopic mass transport and any kind of deformation of the specimen under study appears rational, since electronic excitation under illumination emulates temperature rise above $T_g$. This athermal transformation of the solid viscosity to values corresponding to the supercooled liquid, renders relaxation processes fast enough, so as to fall within the observation experimental time.

Delving into the archival literature, one frequently encounters reports which are simply descriptive, postponing a deeper apprehension of the measured effect. Despite that such works enrich the current phenomenology, bearing a hand-waving character they may add confusion towards our understanding of photoinduced effects. Indeed, several studies have been highly unmethodical reporting the occurrence of some type of photoinduced effect for a particular glass or amorphous film composition, eluding the temptation to consider important aspects of the effect in a more systematic approach. Besides, in several cases, the prosed "structural models" are simple handmade drawings, being highly irrelevant to real atomic arrangements of the structure in question. Understanding the microscopic origin of photoinduced phenomena is a tantalizing task. Although proposed universalities, established for a large number of materials explored, do indeed hold for few effects, it is not uncommon that exceptions to the rule are also frequently met in several cases. The fact that the very same amorphous chalcogenide prepared by a different method, frequently exhibits dissimilar response under illumination, has been an extra source of perplexity in the field.

Research in amorphous chalcogenides is systematically diminishing worldwide over the last years, as other (nano)materials with (alleged) potential for applications emerge at a frantic pace. However, the interest in chalcogenide research will be continued since novel aspects of known effects will keep drawing our attention. Closing this review, we will refer to few such issues that may deserve further exploration as they could either provide better understanding of known effects or result in fascinating new findings.

Computer simulations, such as *ab initio* molecular dynamics, being among the most accurate approaches, have provided instructive results on the structure of various non-crystalline chalcogenides, with emphasis on the last decade on the phase-change mechanisms in Ge-Sb-Te alloys [145, 146]. However, what is perhaps surprising is the lack of systematic computer simulations to guide and augment experiments on interpreting *photoinduced effects*. However, this is not unreasonable from the viewpoint that simulating the photo-electronic excitation process and the ensuing relaxation pathways, is a formidable task. Among the very few articles dealing with theoretical/simulation aspects we should refer to the modelling of the structure of $As_2S_3$ clusters and their electronically excited states using *ab initio* quantum chemical calculations [147]. The existence of localized electronic states of amorphous semiconducting chalcogenides facilitates such an approach because the PiS changes can describe by the local nature of excited electrons at a good accuracy. More recently, modeling $As_2S_3$ as an incompressible viscous fluid, mass transport and surface morphological changes were studied as the result of an optically induced pressure acting on induced dipoles [148]. Simulations towards this direction are expected to shed light on the origin of PiS changes and provide a quantitative means for a priori predictions.

It might seem surprising that photoplastic effects were first reported and methodically explored for crystals (since 1957), as briefly surveyed in Section. 12.6.1.1. A very recent report has revived the interest in such studies. While the role of light is considered to be positive – in the sense that it facilitates structural changes at the atomic level – Oshima *et al*. [149] presented findings that darkness can induce extraordinary plasticity of bulky ZnS crystals; hence, unsettling the widely accepted perception that inorganic semiconductors fail in a brittle manner under stress. The unexpected high plastic behavior of crystalline ZnS, withstanding deformation strain of ~45%, observed at room temperature is the result of complete darkness, while under white or UV light irradiation the crystals were fractured at few % of strain, as shown in Fig. 19. The role of dislocation motion is key to understanding the observed effects. The plastic deformation is accompanied by dramatic decrease of the optical bandgap from 3.52 for the undeformed crystal to 2.92 for the crystal deformed up to 35% plastic strain. Questing similar effects in bulk chalcogenide glasses would undoubtedly be a stimulating topic.

Truly athermal, photoinduced phenomena have pervasively fascinated researchers working on non-crystalline chalcogenides as they form the ground of a number of potential applications in photonics and optoelectronics. On another front, it has always been challenging to demonstrate the purely athermal nature of the effect under study as irradiation can inevitably transfer some undesired amount of heat to the exposed material. Reversible, optically-driven athermal phase changes between a crystal and its disordered state have been demonstrated in a few cases. Recent findings [143] of an all-optical-based mechanism eliciting the reversible transition: glassy(amorphous) ↔ (metastable) crystalline ↔ reamorphized solid, are of particular interest and certainly deserve more systematic studies. The interest particularly arises from our poor, yet, understanding of the second step of the transition stated above. Even scarcer are athermal light induced amorphous ↔ amorphous transformations, particularly when they are accompanied by strong changes in optical properties and can be induced in successive cycles [109]. The nature of polyamorphic transitions between two structurally distinct meta-stable amorphous phases driven solely by light, is a highly unexplored theme.

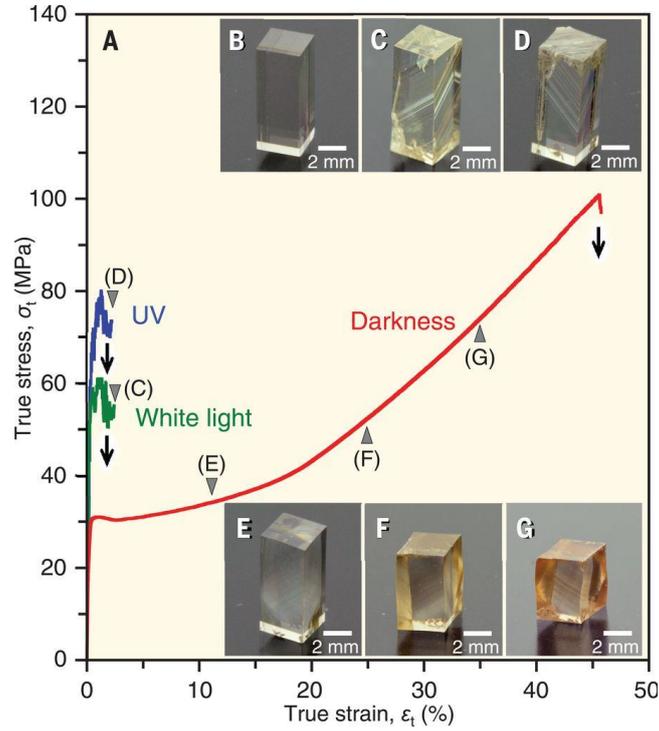

**Fig. 19.** (A) Stress-strain curves of ZnS single crystals. (B) An as-grown crystal. (C), (D) ZnS crystals deformed under white and UV light. (E), (F), (G) Crystals deformed at certain strains as shown in the solid curve in (A). Reprinted from Ref. 149. Copyright © 2018 The Authors, some rights reserved; exclusive licensee American Association for the Advancement of Science.

The emerging field of 2D crystals such as transition metal dichalcogenides (TMDChs) of the form $MX_2$ (M: Mo, W, …, X: S, Se, Te,) might open up new directions and possibilities for the demonstration of new potential functionalities arising from photo-stimulated effects. Despite that the demonstration of photoinduced effects in TMDChs materials is highly challenging due to the very poor glass-forming ability these materials exhibit, this is still an open field of research pervaded with opportunities and prospects.

Benefiting from micro-fabrication capabilities arising from photoplastic effects, amorphous chalcogenides could apparently meet a number of applications in photonics. Counter to conventional heat methods which can be used to plastically deform or mold glassy/amorphous solids, the employment of light offers certain advantages. The avoidance of heat conduction and the precision targeting to the desired area for molding material's' shape, are amongst the most prominent. Even if the transition of such effects to commercial applications still seems remote or elusive, exploring the origin of photoinduced effects is of utmost interest in its own right.

Undeniably, non-crystalline chalcogenides do not possess the exclusive privilege of being the sole light-responsive materials. However, they settle indeed atop this pyramid being the class of materials exhibiting the richest variety of photoinduced effects in comparison to other types. The systematic studies and the current understanding in this field of research might unlock the potential for clarifying similar effects in less explored areas.

**Acknowledgments**

I want to thank a number of individuals who have contributed to the work presented in this chapter. At first, I would like to express my gratitude to Prof. G. N. Papatheodorou (my Ph.D. supervisor) for the guidance, advice and encouragement he provides constantly since 1991. His acquaintance with the (late now) Prof. H. Fritzsche (U. Chicago) was the trigger which initiated my exploration of photoplastic effects in chalcogenides almost 18 years ago. Since then, I have been cooperating on this research topic with a number a number of collaborators (Dr. D. Th. Kastrissios, Dr. K. S. Andrikopoulos, Dr. M. Kalyva, Dr. A. Siokou, Dr. F. Kyrazis, Ms. V. Benekou, Dr. J. Orava, and Prof. T. Wagner) whom I wish to thank. Illuminating discussions with Dr. R. O. Jones and A. V. Kolobov on specific aspects of this chapter, as well as the discussion and the assistance of Dr. M. L. Trunov with the Russian literature are highly appreciated. Finally, I want to thank my Ph.D. student Ms. A. Antonelou for her valuable help in sorting the references of the present document and the ICE/HT librarian Ms. M. Perivolari for managing the copyrighted material presented herein.